\documentclass[journal]{IEEEtran}
\usepackage{amsmath}
\usepackage{amsmath,amssymb}
\usepackage{stmaryrd}
\usepackage{graphicx,graphics,color,psfrag}
\usepackage{cite,balance}
 \usepackage{subfigure}
 \usepackage{booktabs}
\usepackage{caption}   
\allowdisplaybreaks
\usepackage{algorithm}
\usepackage{bm}
\usepackage{eufrak}
\usepackage{url}
\usepackage[english]{babel}
\usepackage{algpseudocode}
\usepackage{multirow}
\usepackage{enumerate}
\usepackage{cases}

\begin{document}
 \pdfminorversion=4
\graphicspath{{figure/}}
\title{Towards Overfitting Avoidance: Tuning-free Tensor-aided Multi-user  Channel Estimation for 3D Massive MIMO Communications}
\author{Lei Cheng  and Qingjiang Shi \thanks{L. Cheng is with the Shenzhen Research Institute of Big Data, Shenzhen, Guangdong, P. R. China (e-mail: leicheng@sribd.cn).} \thanks{Q. Shi is with the School of Software Engineering at Tongji University, Shanghai, 201804, China. He is also with the Shenzhen Research Institute of Big Data, Shenzhen, 518172, China. Email: shiqj@tongji.edu.cn.}}
\maketitle

\begin{abstract}
Channel estimation has long been deemed as one of the most critical problems in three-dimensional (3D) massive multiple-input multiple-output (MIMO), which is recognized as the leading technology that enables 3D spatial signal processing in the fifth-generation (5G) wireless communications and beyond. Recently, by exploring the angular channel model and tensor decompositions, the accuracy of single-user channel estimation for 3D massive MIMO communications has been significantly improved given a limited number of pilot signals. However, these existing approaches cannot be straightforwardly extended to the multi-user channel estimation task, where the base station (BS) aims at acquiring the channels of multiple users at the same time. The difficulty is that the  coupling among multiple users' channels makes the channel estimation deviate from widely-used tensor decompositions. It gives a non-standard tensor decomposition format that has not been well tackled. To overcome this challenge, besides directly fitting the new tensor model for channel estimation to the wireless data via block coordinate descent (BCD) method, which is prone to the overfitting of noises or requires regularization parameter tuning, we further propose a novel tuning-free channel estimation algorithm that can automatically control the channel model complexity and thus  effectively avoid the overfitting. Numerical results are presented to demonstrate the excellent performance of the proposed algorithm in terms of both estimation accuracy and overfitting avoidance. 
\end{abstract}

\begin{IEEEkeywords}
Joint model-and-data-driven wireless communications, 3D massive MIMO, channel estimation, tensor methods, tuning-free.
\end{IEEEkeywords}

\section{Introduction}
In recent years, massive multiple-input multiple-output (MIMO) has gradually evolved from being a theoretical concept to a leading practical technology for the next generation wireless communications \cite{Larsson1, Larsson2}. To further embrace the forthcoming era of Internet of Things (IoT), it calls for advanced three-dimensional (3D) spatial signal processing techniques  (e.g., 3D beamforming) \cite{3DB1, 3DB2} to allow high-quality communications among multiple users (including unmanned aerial vehicles (UAVs) in the sky \cite{UAV1} and unmanned ground vehicles (UGVs) on roads \cite{UGV1}). To achieve this, rather than mechanically tilting conventional antenna arrays, 3D massive MIMO, in which the BS is with a 3D antenna array, has emerged as an enabling technique to broaden the scope of  BS \cite{3DMIMO1,3DMIMO2,3DMIMO3}. However, its promise can only be fulfilled when accurate channel state information (CSI) of multiple users is available at the BS.

As one of the most critical problems in wireless communications, channel estimation has been continuously studied for many decades, synergizing nearly all the ideas of signal processing methods including optimizations \cite{CE1}, statistics \cite{CE2}, algebras \cite{CE3}, and machine learning \cite{CE4}. Its ultimate goal is to estimate wireless channels as accurate as possible given a limited number of pilot signals, while its challenges vary significantly under different channel models and wireless systems including  millimeter-wave massive MIMO \cite{R1}, IoT \cite{R2} and machine-type communications (MTC) \cite{R3}. That is why ``no free lunch theorem" \cite{ML} in machine learning  also holds in channel estimation research, in the sense that there is no panacea that suits every wireless scenario. In particular, for 3D massive MIMO communications, it is widely recognized that there is a unique challenge in exploiting the inherent 3D spatial structure inside the channel coefficients \cite{3DMIMOCE1, 3DMIMOCE2, CLJSTSP}, and thus needs tailored algorithm designs.

To reveal the underlying 3D spatial structure, an emerging trend is to leverage the angular channel model, which has been validated by real-world measurements \cite{3DMIMO2,3DMIMO3}. In this model, the channel coefficient is modelled as the summation of different propagation paths, each of which is specified by angle parameters and fading parameters. Since this channel model mimics the signal propagations in the physical world, its parameters are with clear interpretations. On the other hand, the mathematical form of the angular channel model shares a lot of similarities with the models in array signal processing, and thus has triggered tremendous research progress on massive MIMO channel estimation from an array signal processing perspective \cite{JSTSP_intro}. In particular, many fundamental ideas in array signal processing, including discrete Fourier transformation (DFT) \cite{DFT}, multiple signal classification (MUSIC) \cite{MUSIC}, and estimation of signal parameters via rotational invariance technique (ESPRIT) \cite{ESPRIT}, have tapped into the algorithm design of massive MIMO channel estimation and brought significant performance improvement \cite{Arr1, Arr2, Arr3}. Furthermore, these inspiring ideas have integrated with advanced tensor methods to achieve more accurate 3D massive MIMO channel estimation even with limited pilot signals \cite{CLTSP1, CLJSTSP, CLSPAWC,  TensorCE1, TensorCE2}.

However, previous works on tensor-aided 3D massive MIMO communications  \cite{CLJSTSP, CLSPAWC,  CLTSP1, TensorCE1, TensorCE2} mainly investigated single-user channel estimation. By equivalently formulating channel estimation problems as standard tensor decompositions, a vast number of off-the-shelf tensor decomposition tools can be utilized \cite{tensor_survey}. One might contemplate the straightforward extension of existing works to the scenario where the BS estimates multiple users' channels simultaneously. Unfortunately, the coupling among different users' channels make the problem formulation deviate from widely-used tensor decompositions. Instead, it gives a non-standard  tensor decomposition format that has not been well tackled. This unique challenge requires a novel tensor-aided multi-user channel estimation algorithm design for 3D massive MIMO communications.

The most straightforward approach is to fit the new tensor decomposition model to the observation data via solving an optimization problem. In particular,  with the widely adopted least-squares (LS) model fitting criterion, it can be shown that the optimization problem enjoys a block multi-convex property \cite{bcd}, in the sense that although the original problem is not convex, after fixing other variables other than one variable, the remaining problem is convex. It motivates the leverage of block coordinate descent (BCD) method \cite{large_scale1} to solve the model fitting problem. This approach can be interpreted as a maximum-likelihood (ML) approach under the assumption that the signals are corrupted by additive white Gaussian noises (AWGNs) \cite{ML}. From the viewpoint of machine learning, it is well known that the ML solution is prone to the overfitting of noises if the model complexity is not set correctly \cite{ML}. In the angular channel model, the model complexity is determined by the number of independent  propagation paths, which however is unknown in practice \cite{3DMIMO2, 3DMIMO3}. To mitigate the overfitting, a  typical method is to introduce an additive regularization term that penalizes complicated channel models \cite{ML}. However, for the best channel estimation performance, this approach requires tuning the regularization parameters carefully to balance the data fitting and the model complexity control, which inevitably consumes enormous computation resources. Therefore, in this paper, we aim at answering the following question: \emph{could we develop a tuning-free channel estimation algorithm that can automatically learn the optimal channel model complexity from the wireless data?} 

This question invites a data-driven approach to the wireless research, in order to let the wireless data tell its desired  channel model complexity. This goal just coincides with the fundamental philosophy of Bayesian methods\cite{beal}. In particular, the Bayesian Occam Razor principle states that the multiple integrations in the Bayes rule will automatically drive the inferred model to the simplest one that can still explain the data well. This has enabled tuning-free algorithm designs for Bayesian neural network \cite{Mackay},  sparse Bayesian learning \cite{Tipping}, and more recently Bayesian structured tensor decompositions \cite{CL2, CL5}. Its great success in automatic model complexity control inspires us to rethink the multi-user 3D massive MIMO channel estimation problem from a Bayesian perspective. In particular, by establishing the probabilistic model and designing the efficient inference algorithm, in this paper, we propose a novel tuning-free tensor-aided multi-user channel estimation algorithm for 3D massive MIMO communications. Numerical results have corroborated its excellent performance in terms of both channel estimation accuracy and overfitting avoidance. 

The remainder of this paper is organized as follows. In Section II, after introducing the system model, the multi-user channel estimation problem is formulated as a non-standard tensor decomposition problem. To fit the new tensor model to the wireless data, a BCD-based method is briefly introduced in Section III, which however is prone to the overfitting of noises. To avoid the overfitting via a tuning-free approach, a novel algorithm based on  Bayesian modelling and inference is proposed in Section IV.  Simulation results are presented in Section V to show the effectiveness of the proposed algorithm. Finally,  conclusions are drawn in Section VI.

%
%
%
%
%
%
%

\textbf{Notation}: Boldface lowercase and uppercase letters will be used for vectors and matrices, respectively.
Tensors are written as calligraphic letters. $\mathbb E [~\cdot~] $ denotes the expectation of its argument and $j \triangleq \sqrt{-1}$. Superscripts $T$, $*$ and $H$ denote transpose, conjugate and Hermitian, respectively. $\boldsymbol A^{-1}$ denotes the inverse of a matrix $\boldsymbol A$. The operator $\textrm{Tr}\left( {\boldsymbol{A}} \right)$ denotes the trace of a matrix $\boldsymbol{A}$. $\bigparallel \cdot {\bigparallel}_F$ represents the Frobenius norm of the argument.  $\mathcal {CN} (\boldsymbol x | \boldsymbol u, \boldsymbol R)$ stands for the probability density function of a   circularly-symmetric complex Gaussian vector $\boldsymbol x$ with mean $\boldsymbol u$ and covariance matrix $\boldsymbol R$. The operator $\mathfrak{Re}\{ \cdot\}$ represents the real part of the argument. The symbol $\propto$ represents a linear scalar relationship between two real-valued functions. The $N \times N $ diagonal matrix with diagonal elements $a_1$ through $a_N$ is represented as $\mathrm{diag} \{ a_1, a_2,...,a_N\}$, while $\boldsymbol I_{M}$ represents the $M \times M$ identity matrix. The $(i,j)^{th} $ element, the $i^{th}$ row, and the $j^{th}$ column of a matrix $\boldsymbol A$ are represented by   $\boldsymbol{A}_{i,j}$,  $\boldsymbol{A}_{i,:}$ and $\boldsymbol{A}_{:,j}$, respectively.

\section{System Model And Problem Formulation: When Angular Channel Model Meets Multi-user Massive MIMO}

Consider a massive MIMO system where the BS is equipped with a 3D uniform cuboid antenna array (UCA), as shown in Figure 1, and each user is equipped with a single antenna.  Let $M$ and $N$ denote the number of antennas at the BS and the number of users, respectively. In the BS, with the first antenna assumed to be the origin of the coordinate system,  the number of antennas in the x-direction, y-direction and z-direction are $I_1$, $I_2$ and $I_3$, respectively (i.e., $M = I_1I_2I_3$). Obviously, the UCA includes the uniform rectangular array (URA)  and the uniform linear array (ULA) as its special cases by setting some of $\{I_1, I_2, I_3\}$ to be one.  

In this paper, we consider the uplink transmission where all the users simultaneously transmit their pilot signals to the BS through narrow-band non-line-of-sight (NLOS) channels\footnote{The discussions on incorporating the LOS path are presented in \emph{Remark 1} (at the end of Section IV. A)}. Each user is assigned a unique pilot sequence $\boldsymbol s_n=[s_n(1),...,s_n(L)]^T$ with length $L$, which is assumed to be smaller than the channel coherence length. The channel state information (CSI) from the $n^{th}$ user to the $m^{th}$ antenna at the BS is modeled as a complex coefficient $h_m^n$. Then, the received discrete-time complex baseband signal at the BS can be modeled as
\begin{align}
\boldsymbol Y &= \sum_{n=1}^N  \boldsymbol s_n \boldsymbol h_n^T + \boldsymbol W= \boldsymbol S \boldsymbol H + \boldsymbol W,
\end{align}
where vector $\boldsymbol h_n = [h_1^n, h_2^n, ..., h_M^n]^T$ collects channel coefficients for the $n^{th}$ user, and each element $w_{l,m}$ in the noise matrix $\boldsymbol W$  denotes the additive white Gaussian noise (AWGN) at the BS, i.e., $w_{l,m} \sim \mathcal {CN}( 0 , \beta^{-1})$ is spatially and temporally independent. Pilot matrix $\boldsymbol S \in \mathbb C^{L\times N}$ is with the $n^{th}$ column being $\boldsymbol s_n$, and channel matrix  $\boldsymbol H \in \mathbb C^{N\times M}$ is with the $n^{th}$ row being $\boldsymbol h_n^T$.

The goal of multi-user channel estimation is to estimate the channel matrix $\boldsymbol H$ from the received data $\boldsymbol Y$ at the BS with the help of the pilot matrix $\boldsymbol S$. From data model (1), a standard least-squares (LS) solution can be obtained immediately:
\begin{align}
\hat {\boldsymbol H}^{\mathrm{LS}} = (\boldsymbol S^H \boldsymbol S)^{-1} \boldsymbol S^H \boldsymbol Y.
\end{align}
When using the LS estimator, since no prior information is incorporated, it is well known that the estimation accuracy heavily relies on the pilot length $L$ \cite{Kay}. That is, to ensure accurate channel estimation, long pilot sequences are required to be transmitted at user sides, which however will   consume invaluable spectral resources. This is not desirable in practical massive MIMO systems, and thus calls for alternative solutions that can significantly improve the  accuracy of channel estimation  even with limited pilots. 

\begin{figure}[!t]
\includegraphics[width= 4 in]{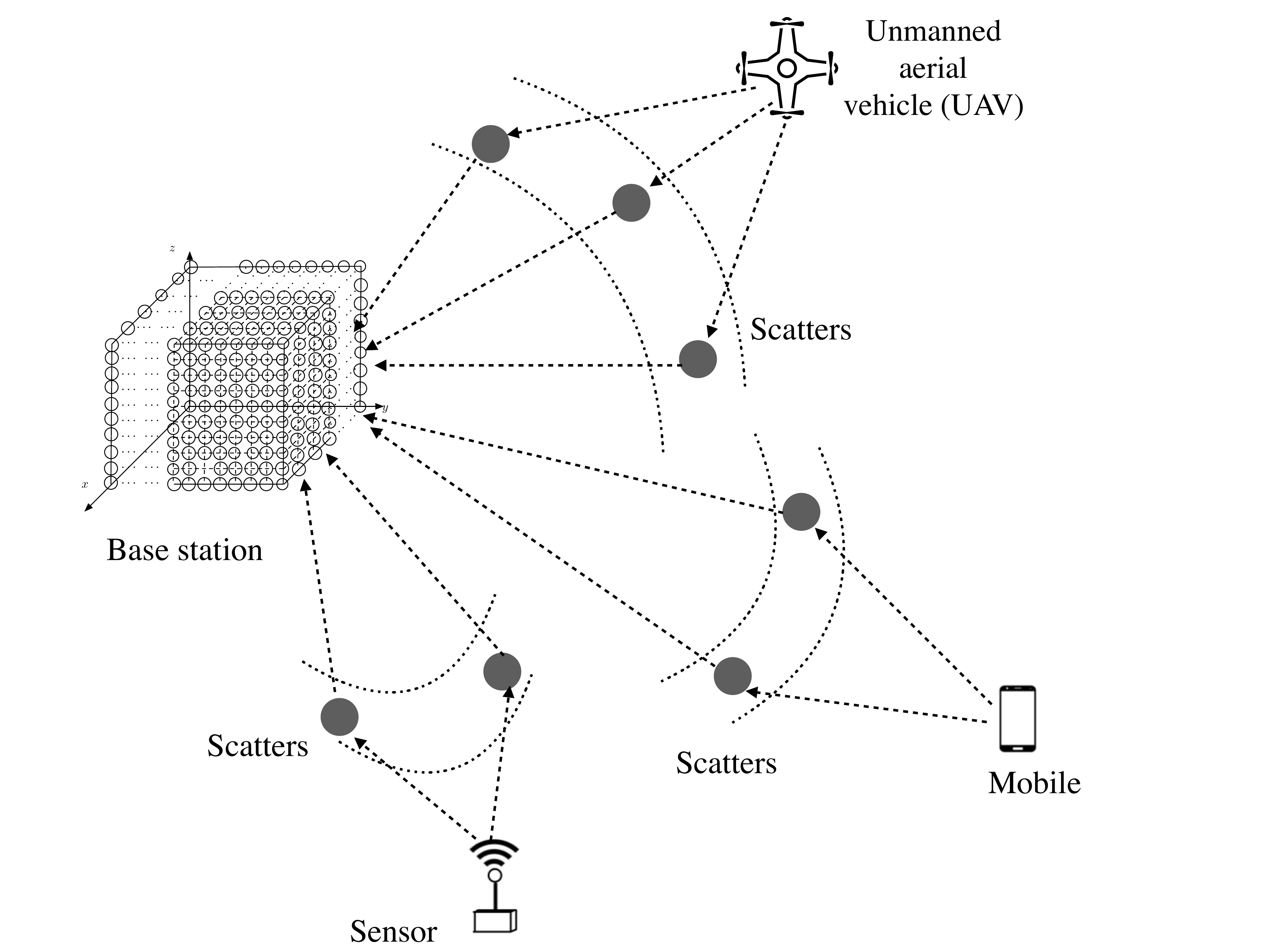}
\caption{A massive MIMO system where the base station (BS) is equipped with a three dimensional (3D) uniform cuboid antenna array (UCA). Under both spatial and frequency narrow-band assumption, the relative delays among different non-line-of-sight (NLOS) propagation paths and the effect of different subcarriers are assumed to be negligible \cite{A3}.}
\label{fig_topology}
\end{figure}


To achieve this, recent research works have repeatedly shown the glimmers of hope from channel model structure exploitation. In particular, a vast amount of research works \cite{JSTSP_intro, CE3, Arr1, Arr2, Arr3, CLJSTSP, CLSPAWC,  CLTSP1, TensorCE1, TensorCE2}  have shown the effectiveness of the angular channel model, which has been validated by real-world measurements \cite{3DMIMO2, 3DMIMO3}. Not only does it depict the signal propagations in the physical word via angle parameters and fading parameters, it also bridges the design of massive MIMO systems and array signal processing techniques. More specifically, it assumes that for the $n^{th}$ user, the channel model consists of $R^n$ propagation paths, each of which is determined by path gain $\xi_{r^n}$,  elevation angle $\theta_{r^n}$ and azimuth angle $\phi_{r^n}$, i.e., 
\begin{align}
&h_{x_m, y_m, z_m}^n  = \sum_{r=1}^{R^n} \xi_{r^n} \mathrm{exp} \Big\{j\frac{2\pi }{\lambda_c}  \big[x_m \sin \theta_{r^n} \cos \phi_{r^n} \nonumber \\
&  +y_m \sin \theta_{r^n} \sin \phi_{r^n}  + z_m \cos \theta_{r^n}\big]\Big\},
\end{align}
where $\lambda_c$ is the wavelength of the carrier signal and $(x_m,y_m,z_m)$ is the coordinate of the $m^{th}$ antenna.  Notice that in (3), under both spatial and frequency narrow-band assumption, the relative delays among different propagation paths and the effect of different subcarriers are assumed to be negligible \cite{A3}. Although the channel model (3) shares a lot of  similarities with the array signal processing model [20], [21], there is a slight difference. Due to the block-fading assumption, the path gain $\xi_{r^n}$ is assumed to be unchanged during the channel estimation. In contrast, in most array signal processing applications [20], [21], the source signals are assumed to be time-varying.

Using the angular channel model (3),  instead of directly estimating the  channel matrix $\boldsymbol H$ with $MN$ unknown parameters, one could estimate the model parameters $\{\{\xi_{r^n}, \theta_{r^n},  \phi_{r^n} \}_{r=1}^{R^n} \}_{n=1}^N$ and then reconstruct the channel coefficients. By this approach, only $\sum_{n=1}^N 3R^n$ unknown parameters need to be estimated. Since the path number $R^n$ is usually much smaller than the antenna number $M$, the adoption of the angular channel model significantly reduces the number of unknown parameters, and thus allows more accurate channel estimation.

However, estimating these unknown parameters $\{\{\xi_{r^n}, \theta_{r^n},  \phi_{r^n} \}_{r=1}^{R^n} \}_{n=1}^N$ from the observation data $\boldsymbol Y$ is quite challenging, since they are nonlinearly coupled in the channel model (3). In particular, motivated by the AWGN assumption, the following LS-based optimization problem can be formulated: 
\begin{align}
&\min_{\{\{\xi_{r^n}, \theta_{r^n},  \phi_{r^n} \}_{r=1}^{R^n} \}_{n=1}^N}  \sum_{l=1}^L \sum_{m=1}^M \bigparallel \boldsymbol Y_{l,m} - \sum_{n=1}^N s_n(l) \nonumber \\
&~~~~~~~~\times \sum_{r=1}^{R^n} \xi_{r^n} \mathrm{exp} \Big\{j\frac{2\pi }{\lambda_c}  \big[x_m \sin \theta_{r^n} \cos \phi_{r^n} \nonumber \\
&~~~~~~~~ +y_m \sin \theta_{r^n} \sin \phi_{r^n}  + z_m \cos \theta_{r^n}\big]\Big\}  {\bigparallel}_F^2. 
\end{align}
Similar optimization problems have been investigated in array signal processing society for many decades  \cite{DFT}, and it is widely agreed that directly optimizing these variables $\{\{\xi_{r^n}, \theta_{r^n},  \phi_{r^n} \}_{r=1}^{R^n} \}_{n=1}^N$  is prohibitively expensive in computations. Instead, subspace methods (e.g, MUSIC \cite{MUSIC} and ESPRIT \cite{ESPRIT}) have come up as the main tools to enable the accurate estimation of these unknown parameters in a computationally efficient manner. Its key idea is to recast the parameter estimation problem as the low dimensional signal subspace recovery problem, for which an array of dimensionality reduction tools (e.g., low-rank matrix decompositions \cite{ML}) are off-the-shelf.  Inspired by this idea and further exploiting the 3D structure of the antenna array at the BS, recent studies leverage low-rank tensor decompositions to achieve better signal subspace recovery and subsequently more accurate channel estimation  \cite{CLTSP1, CLJSTSP, CLSPAWC,  TensorCE1, TensorCE2}.  However, these works are limited to the single-user case, while their extensions to the multi-user scenario are not straightforward.

To see this, following the tensor modelling in previous works   \cite{CLTSP1, CLJSTSP, CLSPAWC,  TensorCE1, TensorCE2}, we re-organize the channel coefficients $\{h_{x_m, y_m, z_m}^{n}\}_{m=1}^M$ into a 3D tensor $\mathcal H^{n} \in \mathbb C^{I_1 \times I_2 \times I_3}$. In particular, let set $S_x$ collect all the antennas' x-axis coordinates $\{x_m\}_{m}^M$, with repeated values eliminated and remaining values sorted via the ascending order, (i.e., $S_x(i_1)$ is the $i_1^{th}$ largest number in $\{x_m\}_{m=1}^M$). Similarly, let $S_y$ and $S_z$ collect ordered coordinates on the y-axis and z-axis, respectively. Then, the channel coefficient $h_{x_m, y_m, z_m}^{n}$ can be equivalently re-indexed as $h_{i_1, i_2, i_3}^{n}$, where the index $(i_1, i_3, i_3)$ satisfies $S_x(i_1) = x_m$, $S_y(i_2) = y_m$ and $S_y(i_3) = z_m$. Using the new indexing scheme, the channel model (3) can be equivalently re-expressed as 
\begin{align}
& h_{i_1, i_2, i_3}^n  = \sum_{r=1}^{R^n} \xi_{r^n} \mathrm{exp} \Big\{j\frac{2\pi }{\lambda_c}  \big[S_x(i_1) \sin \theta_{r^n} \cos \phi_{r^n} \nonumber \\
&  +S_y(i_2) \sin \theta_{r^n} \sin \phi_{r^n}  + S_3(i_3) \cos \theta_{r^n}\big]\Big\}, \nonumber \\
& = \sum_{r=1}^{R^n} \xi_{r^n} \mathrm{exp} (S_x(i_1) u_{r^n} ) \mathrm{exp} (S_y(i_2) v_{r^n} )  \mathrm{exp} (S_z(i_3) p_{r^n} ), 
\end{align}
where $u_{r^n} = j\frac{2\pi}{\lambda_c}  \sin \theta_{r^n} \cos \phi_{r^n}$, $v_{r^n}  = j\frac{2\pi}{\lambda_c}  \sin \theta_{r^n} \sin \phi_{r^n} $ and $p_{r^n}  = j\frac{2\pi}{\lambda_c}    \cos \theta_{r^n}$.  Comparing expression (5) to the definition of tensor canonical polyadic decomposition (CPD) \cite{tensor1}, it is easy to identify that each 3D channel tensor $\mathcal H^n$ follows a rank-$R^n$ tensor CPD format:
\begin{align}
\mathcal H^n & \triangleq  \llbracket \boldsymbol U^{(n)}, \boldsymbol V^{(n)}, \left[\boldsymbol \xi^n\right]^T \diamond \boldsymbol P^{(n)}  \rrbracket \nonumber \\
& = \sum_{r=1}^{R^n} \boldsymbol u^{(n)}_r \circ   \boldsymbol v^{(n)}_r  \circ \boldsymbol p^{(n)}_r  \xi_{r^n},
\end{align}
where  $\boldsymbol U^{(n)} \in \mathbb C^{I_1 \times R^n}$ is with its $(i_1, r)^{th}$ element being $\mathrm{exp} (S_x(i_1) u_{r^n} )$; $\boldsymbol V^{(n)} \in \mathbb C^{I_2 \times R^n}$ is with its $(i_2, r)^{th}$ element being $\mathrm{exp} (S_y(i_2) v_{r^n} )$; and $ \boldsymbol P^{(n)}   \in \mathbb C^{I_3 \times R^n}$ is with its $(i_3, r)^{th}$ element being $\mathrm{exp} (S_z(i_3) p_{r^n} )$. $\boldsymbol u^{(n)}_r$, $\boldsymbol v^{(n)}_r$ and $\boldsymbol p^{(n)}_r $ are the $r^{th}$ columns in matrix $\boldsymbol U^{(n)} $, $\boldsymbol V^{(n)} $ and $\boldsymbol P^{(n)}$, respectively.  Symbol $\circ$ denotes vector outer product, $\diamond$ denotes Khatri-Rao product, and vector $\boldsymbol \xi^n= [\xi_{1^n}, \xi_{2^n}, ..., \xi_{R^n}]^T \in \mathbb C^{R^n \times 1} $. Then, optimization problem (4) can be equivalently formulated as:
\begin{align}
&\min_{\{\{\xi_{r^n}, \theta_{r^n},  \phi_{r^n} \}_{r=1}^{R^n} \}_{n=1}^N}  \sum_{l=1}^L  \bigparallel  \mathcal Y_l - \sum_{n=1}^N s_n(l) \nonumber \\
& ~~~~~~~~~~ \times  \llbracket \boldsymbol U^{(n)}, \boldsymbol V^{(n)}, \left[\boldsymbol \xi^n\right]^T \diamond \boldsymbol P^{(n)}  \rrbracket  {\bigparallel}_F^2.
\end{align}
In (7), $\mathcal Y_l \in \mathbb C^{I_1 \times I_2 \times I_3}$ is a 3D tensor that collects measurements $\{\boldsymbol Y_{l,m}\}_{m=1}^M$ according to $\left[\mathcal Y_l\right]_{i_1,i_2,i_3} = \boldsymbol Y_{l, m^{*}}$ where $x_{m^{*}} = S_x(i_1)$, $y_{m^{*}} = S_y(i_2)$ and $z_{m^{*}} = S_z(i_3)$. At last, inspired by the subspace methods, rather than searching parameters $\{\{\xi_{r^n}, \theta_{r^n},  \phi_{r^n} \}_{r=1}^{R^n} \}_{n=1}^N$ exhaustively, it is viable to firstly estimate the factor matrices $\{ \boldsymbol U^{(n)}, \boldsymbol V^{(n)}, \left[\boldsymbol \xi^n\right]^T \diamond \boldsymbol P^{(n)} \}_{n=1}^N$ from the data $\{ \mathcal Y_l\}_{l=1}^L$ and then reconstruct each channel tensor $\mathcal H^n$ via $\llbracket \boldsymbol U^{(n)}, \boldsymbol V^{(n)}, \left[\boldsymbol \xi^n\right]^T \diamond \boldsymbol P^{(n)}  \rrbracket$.  
For the brevity of notations, let factor matrices $\{ \boldsymbol U^{(n)}, \boldsymbol V^{(n)}, \left[\boldsymbol \xi^n\right]^T \diamond \boldsymbol P^{(n)} \}_{n=1}^N$ simply be denoted by  $\{ \boldsymbol \Xi^{(1),n},\boldsymbol \Xi^{(2),n}, \boldsymbol \Xi^{(3),n} \}_{n=1}^N$.  Then, the channel estimation problem can be formulated as:
\begin{align}
&\min_{\{  \{\boldsymbol \Xi^{(k),n} \}_{k=1}^3 \}_{n=1}^N}  \sum_{l=1}^L  \bigparallel  \mathcal Y_l - \sum_{n=1}^N s_n(l) \nonumber \\
& ~~~~~~~~~~~~~~~~~~~~~~\times  \llbracket \boldsymbol \Xi^{(1),n}, \boldsymbol \Xi^{(2),n}, \boldsymbol \Xi^{(3),n} \rrbracket  {\bigparallel}_F^2.
\end{align}

If $N=1$ (i.e. single-user case), problem (8) can be decomposed into a set of standard tensor CPD problems that enjoy appealing uniqueness property (see Appendix A), for which there are abundant  ``out of the box" algorithms \cite{tensor_survey}. However, when the BS serves multiple users simultaneously, the summand inside the Frobenius norm prohibits the straightforward utilization of standard tensor decomposition tools, and thus make the multi-user channel estimation problem much more challenging than the single-user counterpart   \cite{CLTSP1, CLJSTSP, CLSPAWC,  TensorCE1, TensorCE2}. In particular, when $N >1$, the factor matrices $\{  \{\boldsymbol \Xi^{(k),n} \}_{k=1}^3 \}_{n=1}^N $ are intricately coupled together after expanding the Frobenius norm (as elaborated in Appendix B). This coupling is much more complicated than those appeared in existing single-user channel estimation works  \cite{CLTSP1, CLJSTSP, CLSPAWC,  TensorCE1, TensorCE2}, making their extensions (either using optimizations or Bayesian methods) to the multi-user scenario not straightforward.  This paper makes the first attempt to tackle this challenge.

\section {Direct Fitting via Block Coordinate Descent: How to Avoid Overfitting? }

It is not difficult to show the non-convexity of problem (8), since all the factor matrices  $\{  \{\boldsymbol \Xi^{(k),n} \}_{k=1}^3 \}_{n=1}^N$ are coupled together via multi-linear products. However, a closer inspection could reveal its  appealing block multi-convex property \cite{bcd}, based on which BCD method \cite{large_scale1} can be leveraged.  More specifically, although problem (8) is not convex with respect to $\{  \{\boldsymbol \Xi^{(k),n} \}_{k=1}^3 \}_{n=1}^N$, after fixing all the variables to their latest updates other than a single factor matrix $\boldsymbol \Xi^{(k),n}$, in the iteration $t+1$, the remaining subproblem can be formulated as:
\begin{align}
&\min_{\boldsymbol \Xi^{(k),n}}  \sum_{l=1}^L   \bigparallel \! \left[\mathcal B_{l,  p \neq n} \right]^{\kappa}(k) \!\!-\!\! s_n(l) \boldsymbol  \Xi^{(k),n} \left( \mathop \diamond  \limits_{j=1,j\ne k}^3  \left[\boldsymbol \Xi^{(j),n}\right]^{\kappa} \right)^T \!\! {\bigparallel}_F^2,
\end{align} 
where
\begin{align}
&\left[\mathcal B_{l, p\neq n} \right]^{\kappa} \nonumber \\
&\triangleq  \mathcal Y_l \!\!-\!\! \sum_{p=1, p\neq n}^N s_p(l) \Big \llbracket \!\left[\boldsymbol \Xi^{(1),p}\right]^{\kappa}, \left[\boldsymbol \Xi^{(2),p}\right]^{\kappa}, \left[\boldsymbol \Xi^{(3),p} \right]^{\kappa} \! \Big \rrbracket,
\end{align}
and $\kappa$ denotes the most recent update index, i.e., $\kappa = t+1$ when $j < k$ or $p<n$, and $\kappa = t$ otherwise.   $\left[\mathcal B_{l, n \neq k} \right]^{\kappa}(k) $ is a matrix obtained by unfolding the tensor $\left[\mathcal B_{l, n \neq k} \right]^{\kappa}$ along its $k^{th}$ dimension, and the multiple Khatri-Rao products $\mathop \diamond  \limits_{n=1,n\ne k}^N   {\boldsymbol  A}^{(n)} =  {\boldsymbol  A}^{(N)}  \diamond  {\boldsymbol  A}^{(N-1)} \diamond  \cdots \diamond  {\boldsymbol  A}^{(k+1)} \diamond {\boldsymbol  A}^{(k-1)} \diamond \cdots \diamond  {\boldsymbol  A}^{(1)}$. After checking the positive semi-definiteness of the Hessian matrix, subproblem (9) can be shown to be convex. Then, by setting the derivative of the objective function in (9) to be zero, the closed-form optimal solution can be obtained as follows: 
\begin{align}
& \left[\boldsymbol \Xi^{(k),n}\right]^{t+1} \!\!\!=\! \left[\sum_{l=1}^L \left[\mathcal B_{l, n \neq k} \right]^{\kappa}(k)  s_n(l)^*\left( \mathop \diamond  \limits_{j=1,j\ne k}^3  \left[\boldsymbol \Xi^{(j),n}\right]^{\kappa} \right)^* \right]\nonumber \\
& \!\!\!\times \!\!\! \left[\sum_{l=1}^L |s_n(l)|^2  \left( \mathop \diamond  \limits_{j=1,j\ne k}^3  \left[\boldsymbol \Xi^{(j),n}\right]^{\kappa} \right)^T\!\! \left( \mathop \diamond  \limits_{j=1,j\ne k}^3  \left[\boldsymbol \Xi^{(j),n}\right]^{\kappa} \right)^* \!\right]^{-1}.
\end{align}
Since each subproblem is convex, after iteratively updating each $\left[ \boldsymbol \Xi^{(k),n}\right]^{t+1}$ via (11), the resultant BCD algorithm, which is summarized in {\bf Algorithm 1} at the top of this page,  is guaranteed to converge to a critical point of the objective function of (8) \cite{large_scale1}.

\begin{algorithm}[!t]
    \caption*{\bf Algorithm 1: BCD Based Multi-user Channel Estimation}
\noindent {\bf Initializations:}
Choose path number estimates $\{\hat{R}^n\}_{n=1}^N$ and initial values $ \{\{\left[ \boldsymbol \Xi^{(k),n} \right]^0 \}_{k=1}^3 \}_{n=1}^N$.

\noindent {\bf Iterations:} For the iteration $t+1$ ($t \geq 0$),

\noindent \underline{Update factor matrix   $\left[ \boldsymbol \Xi^{(k),n} \right]^{t+1}$}
\begin{align}
& \left[\boldsymbol \Xi^{(k),n}\right]^{t+1} \!=\! \left[\sum_{l=1}^L \left[\mathcal B_{l, n \neq k} \right]^{\kappa} (k) s_n(l)^*\left( \mathop \diamond  \limits_{j=1,j\ne k}^3  \left[\boldsymbol \Xi^{(j),n}\right]^{\kappa} \right)^* \right]\nonumber \\
& \!\!\!\times \!\!\! \left[\sum_{l=1}^L |s_n(l)|^2  \left( \mathop \diamond  \limits_{j=1,j\ne k}^3  \left[\boldsymbol \Xi^{(j),n}\right]^{\kappa} \right)^T\!\! \left( \mathop \diamond  \limits_{j=1,j\ne k}^3  \left[\boldsymbol \Xi^{(j),n}\right]^{\kappa} \right)^* \!\right]^{-1}, \nonumber
\end{align}
where $\left[\mathcal B_{l, n \neq k} \right]^{\kappa} $ is computed using (10);  $\kappa$ denotes the most recent update index, i.e., $\kappa = t+1$ when $j < k$ or $p<n$, and $\kappa = t$ otherwise.

\noindent {\bf{Until Convergence}}

\noindent \underline{Channel Estimation:} $$\hat{\mathcal H}^n =   \Big \llbracket \left[\boldsymbol \Xi^{(1),n}\right]^{t+1}, \left[\boldsymbol \Xi^{(2),n}\right]^{t+1}, \left[\boldsymbol \Xi^{(3),n} \right]^{t+1} \Big \rrbracket, \forall n.$$
\end{algorithm}

However, to implement {\bf Algorithm 1},  prior knowledge about the path numbers $\{R^n\}_{n=1}^N$ are required, which however is difficult to acquire in practice. On the other hand, as seen in (6),  path number $R^n$ controls the number of rank-1 component in the CPD model, and thus controls the channel model complexity. In \cite{tensor1}, it has been shown that generally $R^n$ is non-deterministic polynomial-time hard (NP-hard) to estimate. With over-estimated path numbers $\{\hat{R}^n\}_{n=1}^N$, (or equivalently too complicated channel models), directly fitting the tensor channel models $\{\mathcal H \}_{n=1}^N$ to the  observation data $\{\mathcal Y_l \}_{l=1}^N$ via {\bf Algorithm 1} will be prone to the overfitting of noises, and thus will cause performance deterioration in channel estimation. To avoid the overfitting, a widely-adopted approach is to introduce an additional regularization term that penalizes the model complexity as follows \cite{ML}: 
\begin{align}
&\min_{\{  \{\boldsymbol \Xi^{(k),n} \}_{k=1}^3 \}_{n=1}^N}  \sum_{l=1}^L  \bigparallel  \mathcal Y_l - \sum_{n=1}^N s_n(l) \nonumber \\
& \times  \llbracket \boldsymbol \Xi^{(1),n}, \boldsymbol \Xi^{(2),n}, \boldsymbol \Xi^{(3),n} \rrbracket  {\bigparallel}_F^2 + \sum_{n=1}^N \sum_{k=1}^3 \gamma_{k}^n g(\boldsymbol \Xi^{(k),n}),
\end{align}
where the regularization function $g(\cdot)$ (e.g., $l_1$ norm and $l_2$ norm) is pre-selected. For the best channel estimation performance,  the regularization parameters $\{\{ \gamma_{k}^n \}_{k=1}^3\}_{n=1}^N$ need to be finely tuned, which is however computationally demanding.  Then, an immediate question is: could we develop a tuning-free approach such that the model complexity can be optimally learnt from the data? This is fundamentally important in achieving overfitting avoidance for channel estimation.

\section {Towards A tuning-free Approach: A Bayesian Perspective}
This question has been partially answered in the research of Bayesian modelling and inference. In the early pioneering works of Mackay \cite{Mackay} and Tipping \cite{Tipping} on Bayesian neural network and relevance vector machine, sparsity-enhancing priors were employed to encode an over-parameterized model. Together with the Bayesian Occam Razor principle, which indicates that Bayesian inference will automatically seek the simplest model that can still explain the data adequately, the inference algorithm will drive redundant model parameters to be all zeros and thus effectively control the model complexity.  This idea has triggered flourishing research on Bayesian compressive sensing \cite{BCS}, sparse Bayesian learning \cite{SBL1}, and more recently Bayesian structured tensor decompositions \cite{CL2, CL5}. However, for the tensor-aided multi-user channel estimation problem in (8), since it does not follow a standard tensor decomposition format,  there is no existing Bayesian solution. Therefore, in this paper, we develop such an algorithm from the first principle of Bayesian methods. 

\subsection {Sparsity-promoting Probabilistic  Modelling}

Firstly, the probabilistic model, which encodes the knowledge of problem (8),  needs to be established. Motivated by the LS cost function in (8) (and equivalently the AWGN assumption in data model (1)), a Gaussian likelihood function is adopted as follows:
\begin{align}
& p(\{\mathcal Y_l\}_{l=1}^L | \{ \{\boldsymbol \Xi^{(k),n}\}_{k=1}^3\}_{n=1}^N) \nonumber\\
& \propto \exp\Bigg\{-\beta \sum_{l=1}^L \bigparallel  \mathcal Y_l - \sum_{n=1}^N s_n(l)  \llbracket \boldsymbol \Xi^{(1),n}, \boldsymbol \Xi^{(2),n}, \boldsymbol \Xi^{(3),n} \rrbracket  {\bigparallel}_F^2\Bigg\},
\end{align}
where $\beta^{-1}$ is the noise power. To reflect the non-informativeness of the noise power, gamma distribution  $p(\beta) = \mathrm{gamma}(\beta | \epsilon, \epsilon)$ with $\epsilon$ being very small (e.g., $10^{-6}$) is employed as its prior.

For factor matrices $ \{\{\boldsymbol \Xi^{(k),n} \}_{k=1}^3\}_{n=1}^N$, since $\{\boldsymbol \Xi^{(k),n} \}_{k=1}^3$ determines the $n^{th}$ user's channel tensor $\mathcal H^n$ and the channels of different users are assumed to be statistically independent, we have $p( \{\{\boldsymbol \Xi^{(k),n} \}_{k=1}^3\}_{n=1}^N) = \prod_{n=1}^N p( \{\boldsymbol \Xi^{(k),n} \}_{k=1}^3)$. In the channel model (6), it can be observed that the channel tensor $\mathcal H^n$ is the summation of $R^n$ rank-1 tensors, each of which is determined by the $r^{th}$ columns in the three factor matrices, i.e.,  $\{ \boldsymbol \Xi^{(k),n}_{:,r} \}_{k=1}^3$. By treating each $\{ \boldsymbol \Xi^{(k),n}_{:,r} \}_{k=1}^3$ as an  independent building block of the channel model, we have $p(\{\boldsymbol \Xi^{(k),n} \}_{k=1}^3) = \prod_{k=1}^3 \prod_{r=1}^{R^n} p( \boldsymbol \Xi^{(k),n}_{:,r})$. Since the exact path number $R^n$ is unknown, an upper bound on its value $\bar{R}^n$ is assumed to give an over-parameterized model. Then, inspired by previous Bayesian sparsity modelling works \cite{Mackay, Tipping}, a sparsity-promoting Gaussian-gamma prior distribution is utilized to model $\{\boldsymbol \Xi^{(k),n}_{:,r}\}_{r=1}^{\bar{R}^n}$. Finally, the sparsity-promoting prior for all the factor matrices is:
\begin{align}
&p( \{\{\boldsymbol \Xi^{(k),n} \}_{k=1}^3\}_{n=1}^N | \{\{\gamma_r^n\}_{r=1}^{\bar{R}^n}\}_{n=1}^N)  \nonumber \\
&= \prod_{n=1}^N p( \{\boldsymbol \Xi^{(k),n} \}_{k=1}^3 | \{\gamma_r^n\}_{r=1}^{\bar{R}^n} )  \nonumber \\
& = \prod_{n=1}^N \prod_{k=1}^3 \prod_{r=1}^{\bar{R}^n} p( \boldsymbol \Xi^{(k),n}_{:,r} |  \gamma_r^n) \nonumber \\
&=  \prod_{n=1}^N \prod_{k=1}^3 \prod_{r=1}^{\bar{R}^n} \mathcal {CN} \left(  \boldsymbol \Xi^{(k),n}_{:,r} | \boldsymbol 0_{I_k}, (\gamma_r^n)^{-1} \boldsymbol I_{I_k} \right),  \\
& p(\{\{\gamma_r^n\}_{r=1}^{\bar{R}^n}\}_{n=1}^N) = \prod_{n=1}^N  \prod_{r=1}^{\bar{R}^n} \mathrm{gamma} (\gamma_r^n | \epsilon, \epsilon),
\end{align}
where $\epsilon$  is a very small number  (e.g., $10^{-6}$) that indicates the non-informativeness of the prior model. Consequently, the proposed probabilistic model is a three-layer Bayes network, as illustrated in Figure 2.

\noindent {\emph{Remark 1:} In (14), due to the NLOS assumption adopted in this paper, there is no need to explicitly model the significant power differences of paths.  If the LOS path is considered (with its power much larger than other paths), order statistics \cite{A2} might be exploited to model this structural information, which is  an interesting future direction to investigate. } 

\begin{figure}[!t]
\centering
\includegraphics[width=3.6 in]{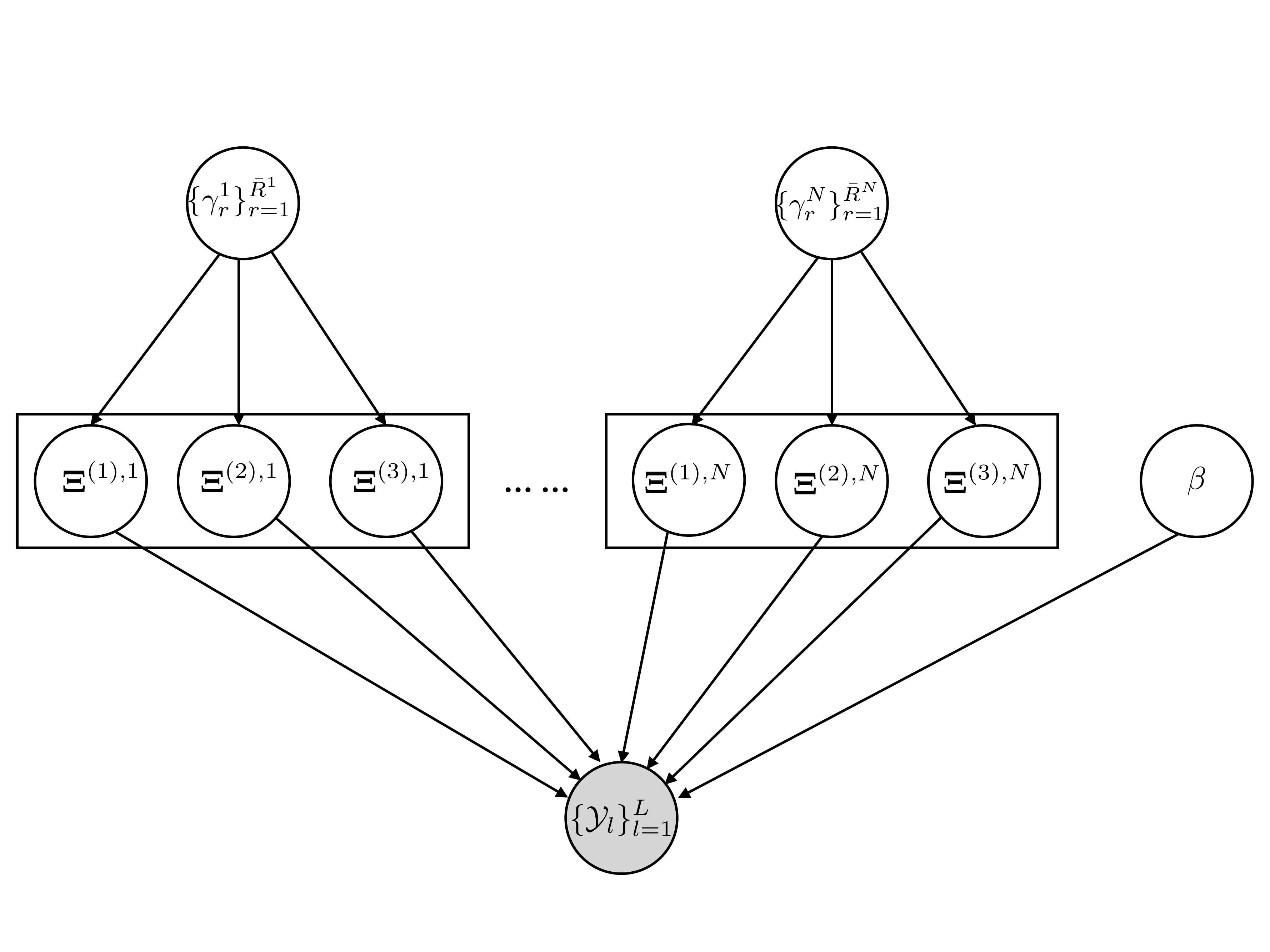}
\caption{Probabilistic model for 3D massive MIMO multi-user channel estimation.}
\label{fig_topology}
\end{figure}

\subsection{Variational Inference: Block Coordinate Descent Over Functional Space}

\begin{table*}[!t]
\centering
\caption{Optimal variational probability density functions}
\begin{tabular}{@{}|l|l|@{}}
\toprule
Variational pdfs  &  Remarks \\ \midrule
$Q^\dagger \left(  \boldsymbol \Xi^{(k),n}\right) = \mathcal{CMN}(\boldsymbol \Xi^{(k),n} |\boldsymbol M^{(k),n}, \boldsymbol I_{I_k}, \boldsymbol \Sigma^{(k),n}), \forall n, k$  &  Circularly-symmetric complex matrix normal distribution \\ & with mean $\boldsymbol M^{(k),n}$ and covariance matrix  $\boldsymbol \Sigma^{(k),n}$ \\ \midrule
$Q^\dagger \left( \gamma_r^n\right)  =  \mathrm{gamma}(\gamma_r^n | a_r^n, b_r^n), \forall n,r$  &  Gamma distribution with shape $a_r^n$ and rate $b_r^n$\\ \midrule
$Q^\dagger \left( \beta \right)  =  \mathrm{gamma}(\beta | c, d) $ & Gamma distribution with shape $c$ and rate $d$ \\ \bottomrule
\end{tabular}
\end{table*}

Let $\boldsymbol \Theta$ be the set that collects all the unknown variables in the probabilistic model, i.e, $\boldsymbol \Theta = \{  \{\{\boldsymbol \Xi^{(k),n} \}_{k=1}^3\}_{n=1}^N,  \{\{\gamma_r^n\}_{r=1}^{\bar{R}^n}\}_{n=1}^N, \beta\}$. The goal of Bayesian inference is to infer the posterior distribution of each unknown variable. Following the Bayes rule, the posterior distribution $p(\boldsymbol \Theta_i | \{ \mathcal Y_l\}_{l=1}^L) =  \int \frac{ p(\boldsymbol \Theta, \{ \mathcal Y_l\}_{l=1}^L)} {\int p(\boldsymbol \Theta , \{ \mathcal Y_l\}_{l=1}^L) d\boldsymbol \Theta } d \boldsymbol \Theta_{j \neq i}$,  where $\boldsymbol \Theta_i$ is part of $\boldsymbol \Theta$ with $\cup_{i=1}^I \boldsymbol \Theta_i = \boldsymbol \Theta$ and $\cap_{i=1}^I \boldsymbol \Theta_i = \O$. However, the intricacy of the probabilistic model does not allow the tractable solution of the multiple integrations involved \cite{ML}. Fortunately, this challenge is not totally new, and is common for modern Bayesian inference tasks such as Bayesian deep learning \cite{stein_vi1} and Bayesian tensor methods \cite{CL2, CL5}. In these works, variational inference (VI) is advocated since it scales well to complicated models with a large number of parameters \cite{VI1, VI2}. In essence, VI recasts the intractable multiple integration problem into a functional optimization problem. In particular, it solves the following problem \cite{VI2}:
\begin{align}
& \min_ {Q(\boldsymbol \Theta)} \mathrm {KL} \big (Q\left( \boldsymbol  \Theta \right) \parallel  p \left(  \boldsymbol  \Theta \mid  \{\mathcal Y_l\}_{l=1}^L \right) \big) \nonumber \\
&~~~~~~~~\triangleq  - \mathbb E_{Q \left( \boldsymbol  \Theta \right) } \left \{ \ln  \frac{p\left(  \boldsymbol  \Theta \mid   \{\mathcal Y_l\}_{l=1}^L \right)}{ Q \left( \boldsymbol  \Theta \right) }  \right\} \nonumber \\ 
&\mathrm{s.t.} ~~ Q(\boldsymbol \Theta) \in \mathcal F,
\end{align}
where $\mathrm  {KL} (\cdot || \cdot)$ denotes  the Kullback-Leibler (KL) divergence between two arguments, and $\mathcal F$ is a pre-selected family of probability density functions (pdfs). The rationale behind problem (16) is: although the exact posterior distribution $p\left(  \boldsymbol  \Theta \mid  \{\mathcal Y_l\}_{l=1}^L \right)$ has no closed-form, we can still seek a tractable variational probability distribution $Q(\boldsymbol \Theta)$ in one family $\mathcal F$ that is the closest to the true posterior distribution $p\left(  \boldsymbol  \Theta \mid  \{\mathcal Y_l\}_{l=1}^L \right)$ in terms of the KL divergence.

The choice of probability distribution family $\mathcal F$ is an art, since it needs to be flexible enough to ensure the freedoms of variational pdfs, while simple enough to enable efficient functional optimization. Mean-field family is such a good choice, as evidenced by a lot of Bayesian inference works \cite{SVI1, ML}. It assumes that $Q(\boldsymbol \Theta) = \prod_{i=1}^I Q(\boldsymbol \Theta_i)$. Then, inspired by its factorized structure, the idea of BCD could be migrated to the functional space. More specifically, after fixing other variational pdfs $\{Q(\boldsymbol \Theta_j)\}_{j \neq i}$ to their latest update results, the pdf $Q(\boldsymbol \Theta_i)$ can be updated via solving the following problem:
\begin{align}
\min_{Q(\boldsymbol \Theta_i)} \int Q(\boldsymbol \Theta_i) \Big(-\mathbb E_{\prod_{j \neq i} Q(\boldsymbol \Theta_j)} \left[ \ln p(\boldsymbol \Theta, \{\mathcal Y_l\}_{l=1}^L)\right] \nonumber \\
 + \ln Q(\boldsymbol \Theta_i) \Big) d{\boldsymbol \Theta_i}.
\end{align}
Using variational calculus, the optimal solution of subproblem (17) can be shown to be \cite{VI1, VI2}:
\begin{align}
Q^\dagger \left(  \boldsymbol \Theta_i\right) =\frac{\exp\left ( \mathbb E_{ \prod_{j \neq i}Q \left(   \boldsymbol \Theta_j\right) } \left [ \ln  {p\left(  \boldsymbol  \Theta , \{\mathcal Y_l\}_{l=1}^L  \right)} \right]  \right)}{\int \exp\left ( \mathbb E_{ \prod_{j \neq i}Q \left(   \boldsymbol \Theta_j\right) } \left [ \ln  p\left(  \boldsymbol  \Theta , \{\mathcal Y_l\}_{l=1}^L  \right) \right]  \right)  d\boldsymbol \Theta_i },
\end{align}
where 
\begin{align}
&\ln  p\left(  \boldsymbol  \Theta , \{\mathcal Y_l\}_{l=1}^L  \right) = \left( \prod_{k=1}^3 I_k L \right) \ln \beta - \beta \sum_{l=1}^L \bigparallel  \mathcal Y_l - \sum_{n=1}^N s_n(l) \nonumber \\
& \times  \llbracket \boldsymbol \Xi^{(1),n}, \boldsymbol \Xi^{(2),n}, \boldsymbol \Xi^{(3),n}  \rrbracket {\bigparallel}_F^2 + (\epsilon -1)\ln \beta - \epsilon \beta  \nonumber \\
&- \sum_{n=1}^N \sum_{k=1}^3 \left[ \mathrm{Tr} \left(\boldsymbol \Xi^{(k),n}  \boldsymbol \Gamma^n  \left[\boldsymbol \Xi^{(k),n}  \right]^H\right)+ I_k \sum_{r=1}^{\bar{R}^n} \ln \gamma_r^n \right] \nonumber \\
& + \sum_{n=1}^N \sum_{r=1}^{\bar{R}^n} \left[ (\epsilon -1) \ln \gamma_r^n - \epsilon \gamma_r^n \right],
\end{align}
and $\boldsymbol \Gamma^n = \mathrm{diag}\{\gamma_1^n, ..., \gamma_{\bar{R}^n}^n \}$.

\subsection{Tuning-free Algorithm Derivation}

After substituting (19) into (18), the optimal solutions $\{Q^\dagger \left(  \boldsymbol \Theta_i\right)\}_{i=1}^I$ can be obtained. Although straightforward as it may seem,  multiple integrations involved in (18) and complicated tensor algebras in (19) both make the derivations technically challenging and tedious. On the other hand, the coupling among different users' channel parameters deviates the algorithm derivations from those developed in related works on single-user channel estimation  \cite{CLTSP1, CLJSTSP, CLSPAWC,  TensorCE1, TensorCE2}. Consequently, it needs much effort to derive the optimal variational pdfs for the multi-user channel estimation problem. To keep the brevity of the main body, we move the lengthy derivations  to Appendix C and only present the final optimal solutions $\{Q^\dagger \left(  \boldsymbol \Theta_i\right)\}_{i=1}^I$  in Table I at the top of this page.

In Table I, the optimal variational pdf for each factor matrix $Q(\boldsymbol \Xi^{(k),n})$ is a  circularly-symmetric complex  matrix normal distribution \cite{MND}, where the covariance matrix 
\begin{align}
\boldsymbol \Sigma^{(k),n} = &  \Bigg[  \sum_{l=1}^L |s_n(l)|^2 \mathbb E\left[ \beta\right] \mathbb E \Big[ \left(\mathop \diamond  \limits_{j=1,j\ne k}^3 \boldsymbol \Xi^{(j),n}\right)^T \nonumber\\
& \times \left(\mathop \diamond  \limits_{j=1,j\ne k}^3 \boldsymbol \Xi^{(j),n}\right)^*  \Big] + \mathbb E \left[ \boldsymbol \Gamma^n \right] \Bigg]^{-1},
\end{align}
and mean matrix
\begin{align}
&\boldsymbol M^{(k),n} \nonumber \\
&=\sum_{l=1}^L   \mathfrak B_{l, p \neq n} (k) s_n(l)^* \mathbb E\left[ \beta \right] \left( \mathop \diamond  \limits_{j=1,j\ne k}^3  \mathbb E\left[\boldsymbol \Xi^{(j),n}\right] \right)^* \boldsymbol \Sigma^{(k),n},
\end{align}
with 
\begin{align}
&\mathfrak B_{l, p \neq n} \nonumber \\
&\triangleq  \mathcal Y_l  \!-\!  \sum_{p=1, p\neq n}^N s_p(l) \Big \llbracket \mathbb E\left[\boldsymbol \Xi^{(1),p}\right],  \mathbb E\left[\boldsymbol \Xi^{(2),p}\right], \mathbb E \left[\boldsymbol \Xi^{(3),p} \right]  \Big \rrbracket.
\end{align}
On the other hand, the optimal variational distributions for each $\gamma_r^n$ and $\beta$ are gamma distributions, with parameters
\begin{align}
& a_r^n = \epsilon + \sum_{k=1}^3 I_k, \\
& b_r^n = \epsilon + \sum_{k=1}^3 \mathbb E \left[  \left[\boldsymbol \Xi^{(k),n}_{:,r} \right]^H \boldsymbol \Xi^{(k),n}_{:,r}  \right], \\
& c = \epsilon + \prod_{k=1}^3 I_k L, \\
& d = \epsilon + \sum_{l=1}^L  \mathbb E \left[ \bigparallel  \mathcal Y_l -  \sum_{n=1}^N s_n(l) \llbracket \boldsymbol \Xi^{(1),n}, \boldsymbol \Xi^{(2),n}, \boldsymbol \Xi^{(3),n} \rrbracket  {\bigparallel}_F^2\right].
\end{align}

In (20)-(26), there are several expectations that need to be computed. Some of them can be directly obtained from their parameters. In particular, 
$\mathbb E [ \boldsymbol \Xi^{(k),n}] = \boldsymbol M^{(k),n}$, $\mathbb E\left[ \gamma_r^n \right] = \frac{a_r^n}{b_r^n}$ and $\mathbb E\left[ \beta \right] = \frac{c}{d}$. Some of them have already computed in previous works \cite{CL2, CLJSTSP}: 
\begin{align}
&\mathbb E \left[  \left[\boldsymbol \Xi^{(k),n}_{:,r} \right]^H \boldsymbol \Xi^{(k),n}_{:,r}  \right] =  \left[\boldsymbol M^{(k),n}_{:,r}\right]^H  \boldsymbol M^{(k),n}_{:,r} + I_k \boldsymbol \Sigma^{(k),n}_{r,r}, \nonumber\\
& \mathbb E \Big[ \left(\mathop \diamond  \limits_{j=1,j\neq k}^3 \boldsymbol \Xi^{(j),n}\right)^T \left(\mathop \diamond  \limits_{j=1,j\ne k}^3 \boldsymbol \Xi^{(j),n}\right)^*  \Big] \\
&=   \mathop \odot \limits_{j=1, j\neq k}^3  \left[  \left[\boldsymbol M^{(j),n} \right]^T \left[\boldsymbol M^{(j),n}\right]^*  + I_j 
\left[\boldsymbol \Sigma^{(j),n}  \right]^*\right],
\end{align}
where  the multiple Hadamard products $\mathop \odot \limits_{n=1,n\ne k}^N   {\boldsymbol  A}^{(n)} =  {\boldsymbol  A}^{(N)}  \odot {\boldsymbol  A}^{(N-1)} \odot  \cdots \diamond  {\boldsymbol  A}^{(k+1)} \odot {\boldsymbol  A}^{(k-1)} \odot \cdots \odot  {\boldsymbol  A}^{(1)}$.
However,  due to the coupling among different users' channel parameters, there is one complicated expectation in (26) that has not been tackled so far. In Appendix D, we show that 
\begin{align}
& \mathbb E \Bigg[ \bigparallel  \mathcal Y_l -  \sum_{n=1}^N s_n(l) \llbracket \boldsymbol \Xi^{(1),n}, \boldsymbol \Xi^{(2),n}, \boldsymbol \Xi^{(3),n} \rrbracket  {\bigparallel}_F^2 \Bigg], \nonumber \\
 = & \bigparallel  \mathcal Y_l {\bigparallel}_F^2 - \mathrm{Tr}\Bigg(2\mathfrak{Re}\Big(\mathcal Y_l(1) \sum_{n=1}^N s_n(l)^*   \left(\mathop \diamond  \limits_{j=2}^3 \boldsymbol M^{(j),n}\right)^* \nonumber\\
 &\times \left[\boldsymbol M^{(1),n}\right]^H \Big) - \sum_{n=1}^N \sum_{p=1, p \neq n}^N  s_n(l) s_p(l)^* \boldsymbol M^{(1),n}  \nonumber \\
 &\times \left(\mathop \diamond  \limits_{j=2}^3 \boldsymbol M^{(j),n}\right)^T \left(\mathop \diamond  \limits_{j=2}^3 \boldsymbol M^{(j),p}\right)^* \left[\boldsymbol M^{(1),p}\right]^H \Bigg)\nonumber\\
 &+ \mathrm{Tr} \Bigg( \sum_{n=1}^N |s_n(l)|^2 \left [  \left[\boldsymbol M^{(1),n}\right]^H  \boldsymbol M^{(1),n} + I_1 \boldsymbol \Sigma^{(1),n} \right]  \nonumber \\
&  \mathop \odot \limits_{k=2}^3 \left [  \left[\boldsymbol M^{(k),n}\right]^H  \boldsymbol M^{(k),n} + I_k \boldsymbol \Sigma^{(k),n} \right]^* \Bigg).
\end{align}

From (20)-(29), it is easy to see that the parameters of each optimal variational pdf $Q^\dagger \left(  \boldsymbol \Theta_i\right)$ rely on the statistics of other variational pdfs $\{Q^\dagger \left(  \boldsymbol \Theta_i\right)\}_{j\neq i}$.  By alternatively updating these variational pdfs, a tuning-free iterative channel estimation algorithm can be summarized in {\bf Algorithm 2} at the top of this page.

\begin{algorithm}[!t]
    \caption*{\bf Algorithm 2: VI Based Multi-user Channel Estimation}
\noindent {\bf Initializations:}
Choose $\bar{R}^n > R^n, \forall n$, initial values $ \{\{ \left[\boldsymbol M^{(k),n}\right]^0 , \left[\boldsymbol \Sigma^{(k),n}\right]^0\}_{k=1}^3\}_{n=1}^N $ and $\epsilon$. Let $\{[a_r^n]^0, [b_r^n]^0, c^0, d^0\} = \epsilon, \forall r,n$.

\noindent {\bf Iterations:} 

For the iteration $t+1$ ($t \geq 0$),

\noindent \underline{Update the parameters of  $Q(\boldsymbol \Xi^{(k),n})^{t+1}$:}
\begin{align}
&\left[\boldsymbol \Sigma^{(k),n}\right]^{t+1} \!\!= \!\! \Bigg[ \! \sum_{l=1}^L |s_n(l)|^2 \frac{c^t}{d^t} \!\! \mathop \odot \limits_{j=1, j\neq k}^3 \!\! \Bigg[ \left[ \boldsymbol M^{(j),n} \right]^{\kappa,T} \!\! \left[\boldsymbol M^{(j),n}  \right]^{\kappa,*} \nonumber\\
& + I_j \left[\boldsymbol \Sigma^{(j),n}\right]^{\kappa,*} \Bigg]  +\mathrm{diag}\left\{ \frac{ [ a^n_1]^t}{[b^n_1]^t}, ..., \frac{[a^n_{\bar{R}^n}]^t}{[b^n_{\bar{R}^n}]^t}\right \} \Bigg]^{-1},
\end{align}
\begin{align}
&\left[\boldsymbol M^{(k),n}\right]^{t+1} \!\!=\!\! \sum_{l=1}^L \Bigg( \mathcal Y_l  \!-\!\!\!  \sum_{p=1, p\neq n}^N s_p(l) \Big \llbracket \left[\boldsymbol M^{(1),p}\right]^{\kappa}, \left[\boldsymbol M^{(2),p}\right]^{\kappa} \!\!\!\!, \nonumber\\
& \left[\boldsymbol M^{(3),p}\right]^{\kappa} \Big \rrbracket\Bigg)(k)s_n(l)^*\frac{c^t}{d^t}  \left( \mathop \diamond  \limits_{j=1,j\ne k}^3  \left[\boldsymbol M^{(j),n}\right]^{\kappa} \right)^* \left[\boldsymbol \Sigma^{(k),n}\right]^{t+1} \!\!\!\!\!\!,
\end{align}
where $\kappa$ denotes the most recent update index, i.e., $\kappa = t+1$ when $j < k$ or $p<n$, and $\kappa = t$ otherwise.

\noindent \underline{Update the parameters of  $Q(\gamma_r^n)^{t+1}$:}
\begin{align}
&\left[a_r^n\right]^{t+1}=   \epsilon + \sum_{k=1}^3 I_k, \\
&\left[b_r^n\right]^{t+1} =  \epsilon +  \sum_{k=1}^3 \left[\boldsymbol M^{(k),n}_{:,r}\right]^{t+1,H}  \left[\boldsymbol M^{(k),n}_{:,r}\right]^{t+1} \nonumber \\
& ~~~~~~~~~~~~~~ + I_k \boldsymbol \left[\boldsymbol \Sigma^{(k),n}_{r,r}\right]^{t+1}.
\end{align}

\noindent \underline{Update the parameters of  $Q(\beta)^{t+1}$:}
\begin{align}
& c^{t+1} = \epsilon + \prod_{k=1}^3 I_k L, \\
& d^{t+1} = \epsilon +  \sum_{l=1}^L \mathfrak f_l^{t+1}, 
\end{align}
where $\mathfrak f_l^{t+1}$ is computed using (29) with  $\{\boldsymbol M^{(k),n}, \boldsymbol \Sigma^{(k),n} \}$ being replaced by 
$\{\left[\boldsymbol M^{(k),n}\right]^{t+1}, \left[\boldsymbol \Sigma^{(k),n}\right]^{t+1}\}, \forall n, k$.

\noindent {\bf{Until Convergence}}

\noindent \underline{Channel Estimation:} $$\hat{\mathcal H}^n =   \Big \llbracket \left[\boldsymbol M^{(1),n}\right]^{t+1}, \left[\boldsymbol M^{(2),n}\right]^{t+1}, \left[\boldsymbol M^{(3),n} \right]^{t+1} \Big \rrbracket, \forall n.$$
\end{algorithm}

\subsection{Intuitive Interpretation of Updating Equations}
\subsubsection{Intuitive interpretation of (20) and (21)} The covariance matrix $\boldsymbol \Sigma^{(k),n}$ of the approximate posterior distribution $Q(\boldsymbol \Xi^{(k),n})$ computed in (20) combines the prior information from $\mathbb E \left[ \boldsymbol \Gamma^n \right]$ and the information from other factor matrices. It is then used as the rotation matrix in the estimation of the factor matrix mean $\boldsymbol M^{(k),n}$ in (21), which takes the linear combination of the observation data and other factor matrices. If there is no prior information $\mathbb E \left[ \boldsymbol \Gamma^n \right]$ and no noise precision estimate $\mathbb E\left[ \beta\right]$, the update equation (21) is very similar to the BCD update in (11), since VI essentially performs BCD steps over the functional space.

\subsubsection{Intuitive interpretation of (23)-(26)} From (23) and (24), it can be seen that $\mathbb E\left[ \gamma_r^n \right] = \frac{a_r^n}{b_r^n}$ is proportional to the inverse of the sum  of  the $r^{th}$ column powers  in all three factor matrices. Therefore, if the $r^{th}$ columns are learnt to be nearly zero, it will give a very large $\mathbb E\left[ \gamma_r^n \right] = \frac{a_r^n}{b_r^n}$, which will further encourage the sparsity of  the $r^{th}$ columns in (21).
On the other hand, it is straightforward to see that (25) is related to the number of observations and (26) approximates the model fitting error.

\subsection{Further Discussions and Insights}
To gain more insights from the proposed algorithm, discussions on its  automatic model complexity control, convergence property, and computational complexity are presented in this subsection.

\subsubsection{Automatic model complexity control} In {\bf Algorithm 2}, although the initial channel model is over-parameterized, there is no need to manually tune any parameter to control the model complexity for overfitting avoidance, since the parameters $\{\{ \frac{c_r^n}{d_r^n} \}_{r=1}^{\bar{R}^n}\}_{n=1}^N$, which are the expectations of $\{\{\gamma_r^n \}_{r=1}^{\bar{R}^n}\}_{n=1}^N$, effectively shrink the values of redundant columns in the factor matrices. In particular, if $\frac{c_r^n}{d_r^n}$ is learnt to be very large, 
they would contribute to the covariance matrix of the factor matrix (as seen in (30)) and then rescale the $r^{th}$ column of the factor matrix  to approach zero values (as seen in (31)).  On the other hand,  parameters $\{\{ \frac{c_r^n}{d_r^n} \}_{r=1}^{\bar{R}^n}\}_{n=1}^N$ will be updated together with other parameters in the algorithm, following the principle of the employed Bayesian framework.

\subsubsection{Convergence property} The algorithm is developed under the framework of mean-field VI, which inherently performs BCD steps over the functional space. Its convergence result has been established in \cite{VI1}. In particular, it has been shown that when the variational pdf is optimized using (18) in each iteration (just as what we have done in this paper), the limit point generated by the BCD steps over the functional space of variational pdfs is guaranteed to be at least a stationary point of the KL divergence in (16)  under the assumption of mean-field family \cite{VI1} .  

\subsubsection{Computational complexity}  In each iteration, the computational complexity of {\bf Algorithm 2} is dominated by the steps of  updating the factor matrices, costing $O(\sum_{n=1}^N \prod_{k=1}^3 3 I_k (\bar{R}^n)^2 + \sum_{n=1}^N\sum_{k=1}^3(\bar{R}^n)^3)$.  The overall complexity is about  $O(q(\sum_{n=1}^N \prod_{k=1}^3 3 I_k (\bar{R}^n)^2 + \sum_{n=1}^N\sum_{k=1}^3(\bar{R}^n)^3))$ where $q$ is the number of iterations required for convergence. Thus it can be seen that the complexity is comparable to that of {\bf Algorithm 1}, in which the computational complexity is $O(q^\prime(\sum_{n=1}^N \prod_{k=1}^3 3 I_k (\hat{R}^n)^2 + \sum_{n=1}^N\sum_{k=1}^3(\hat{R}^n)^3))$ where $q^\prime$ is the number of iterations at convergence.

\section{Numerical Results and Discussions}

In this section, numerical results are presented to assess the channel estimation performance of the proposed tuning-free algorithm (i.e., {\bf Algorithm 2}).  Consider a UCA with $M = 512$ antenna elements, which are deployed in a  3D grid with dimensions $I_1=8$, $I_2=8$, $I_3=8$ and the inter-grid spacing $d_x = d_y = d_z =   \lambda_c/2$. Assume that there are $N = 5$ users simultaneously transmitting signals to the BS. For each user, there are $R^n = 3$ propagation paths with elevation angles randomly selected from $[-\pi/2, \pi/2]$ and  azimuth angles randomly selected from $[-\pi, \pi]$. The pilot length is $L = 10$, and each pilot symbol is sampled from a zero-mean circularly-symmetric complex Gaussian distribution with unit variance. The path gains $\{\xi_{r^n}\}_{r,n}$  are drawn from a zero-mean circularly-symmetric complex Gaussian distribution with unit variance, and without any correlation across $r$ and $n$. The signal-to-noise ratio (SNR) is defined as  $10\log_{10} \left(  \frac{\sum_{l=1}^L \sum_{n=1}^N s_n(l) \bigparallel \llbracket \boldsymbol U^{(n)}, \boldsymbol V^{(n)}, \left[\boldsymbol \xi^n\right]^T \diamond \boldsymbol P^{(n)}  \rrbracket  {\bigparallel}_F^2}  { \parallel \mathcal W \parallel_F^2} \right)$ where $\mathcal W \in \mathbb C^{I_1 \times I_2 \times I_3 \times L }$ is a tensor collecting all the noise samples.   For the proposed tuning-free algorithm,   initial mean  $\left[\boldsymbol M^{(k),n}\right]^0$   for each matrix $\boldsymbol \Xi^{(k),n}$ is drawn from a zero-mean circularly-symmetric complex matrix normal distribution with an  identity covariance matrix, and the  initial covariance matrix is set as  $\left[\boldsymbol \Sigma^{(k),n}\right]^0=  \boldsymbol I_{\bar{R}^n \times \bar{R}^n}$. {The upper bound for channel path $\bar{R}^n = \min\{I_1, I_2, I_3\} = 8$ unless stated otherwise, which is a common practice in Bayesian tensor decompositions [17],[27],[35],[36].}  Each point in the following figures is an average of 100 Monte-Carlo trials.

\subsection{Convergence Property and Automatic Channel Model Complexity Learning}

The convergence behavior of the proposed tuning-free algorithm is shown in Figure 3 under two different SNRs, where the mean-square-error (MSE) of channel estimation $ \frac{1}{MN} \sum_{n=1}^N || \hat{\mathcal H}^n - \mathcal H^n ||_F^2 $ is adopted as the measure. From Figure 3, it can be seen that the MSEs of the proposed algorithm decrease significantly in the first tens of iterations and then gradually converge to stable values. 

\begin{figure}[!t]
\centering
\includegraphics[width=3.5 in]{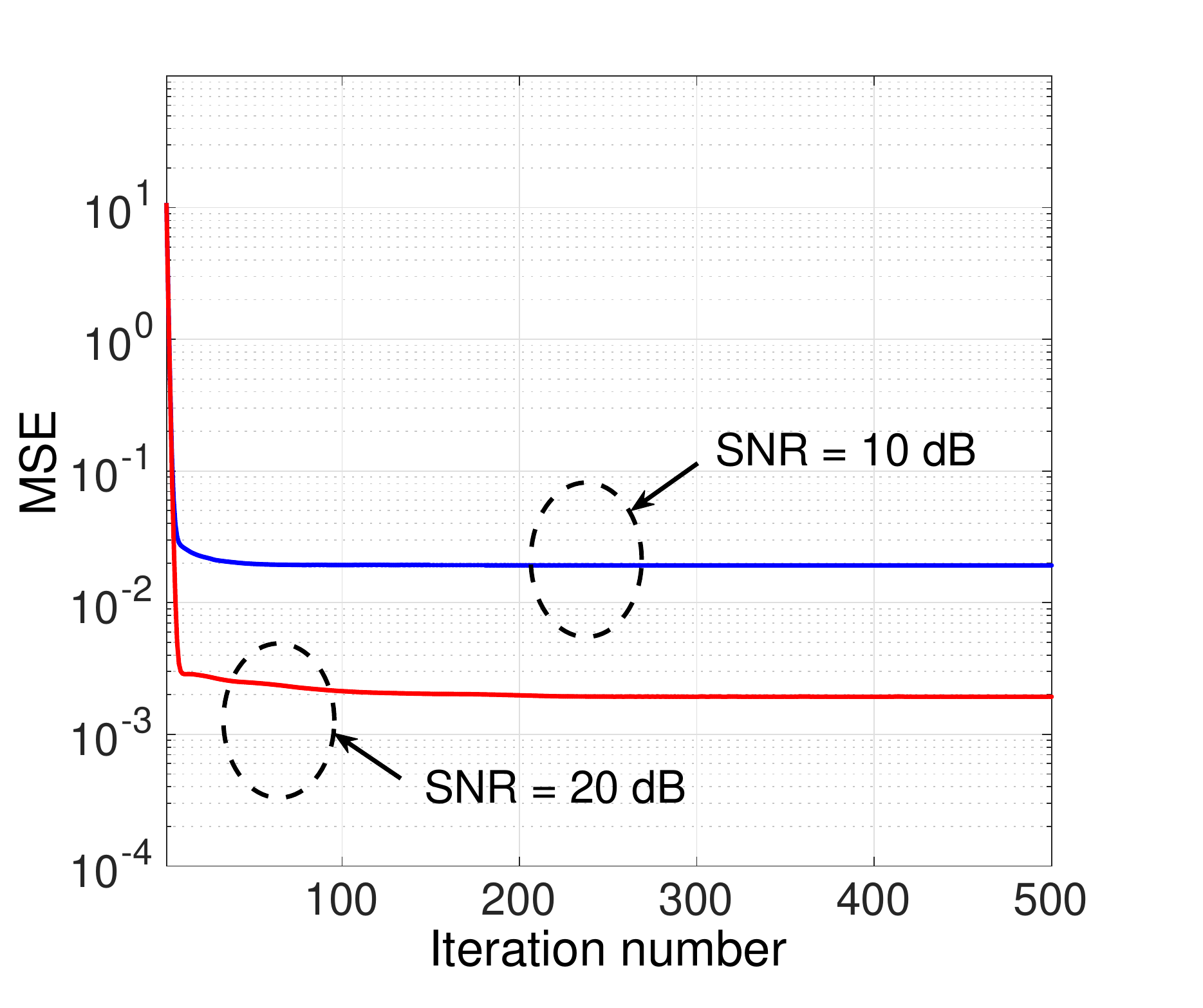}
\caption{The convergence behavior of the proposed algorithm under SNR = 10 dB and SNR = 20 dB ($\bar{R}^n$ = 8, $R^n = 3, L = 10$).}
\label{fig_topology}
\end{figure}

\begin{figure}[!t]
\centering
\includegraphics[width= 3.5 in]{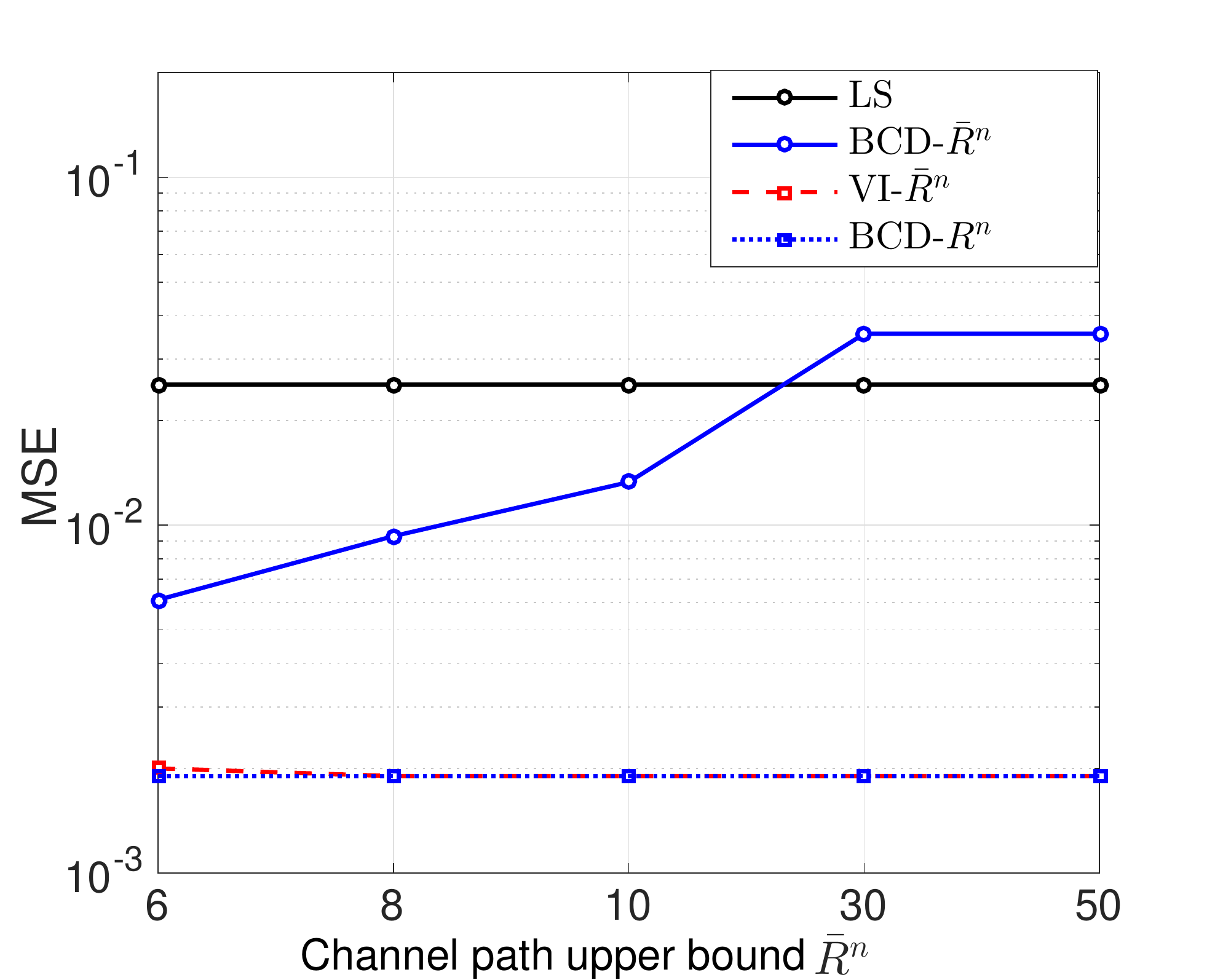}
\caption{Performance of channel estimation versus different channel path upper bound values $\bar{R}^n$ (SNR = 20 dB, $R^n = 3, L = 10$).}
\label{fig_topology}
\end{figure}

To see whether the proposed algorithm is sensitive to the initial upper bound value $\bar{R}^n$, under SNR = 20 dB, the MSEs of the proposed tuning-free algorithm (labeled as VI-$\bar{R}^n$) are presented in Figure 4, in which the MSEs of the LS method (labeled as LS), the BCD method (i.e., {\bf Algorithm 1}) with incorrect path numbers $\{\bar{R}^n\}_{n=1}^N$ (labeled as BCD-$\bar{R}^n$) and the genie-aided BCD method with exact path numbers $\{{R}^n\}_{n=1}^N$ (labeled as BCD-$R^n$) are served as benchmarks. From Figure 4,  it can be seen that the proposed algorithm  with different values of $\bar{R}^n \in \{ 6, 8, 10, 30, 50\}$ shows indistinguishable channel estimation performances to those of the genie-aided BCD-$R^n$ method. Notice that $\bar{R}^n \in \{ 30, 50\}$ is much larger than the true path number (tensor rank) $R^n = 3$.  This shows that with different values of upper bound $\bar{R}^n$, the proposed tuning-free algorithm still can learn the model complexity well and then give accurate channel estimation results. On the other hand, with much larger  $\bar{R}^n \in \{ 30, 50\}$, the BCD-$\bar{R}^n$ algorithm overfits the noises heavily, and even cannot outperform the LS method in channel estimations.

\begin{figure*}[!t]
\centering
\includegraphics[width= 7  in]{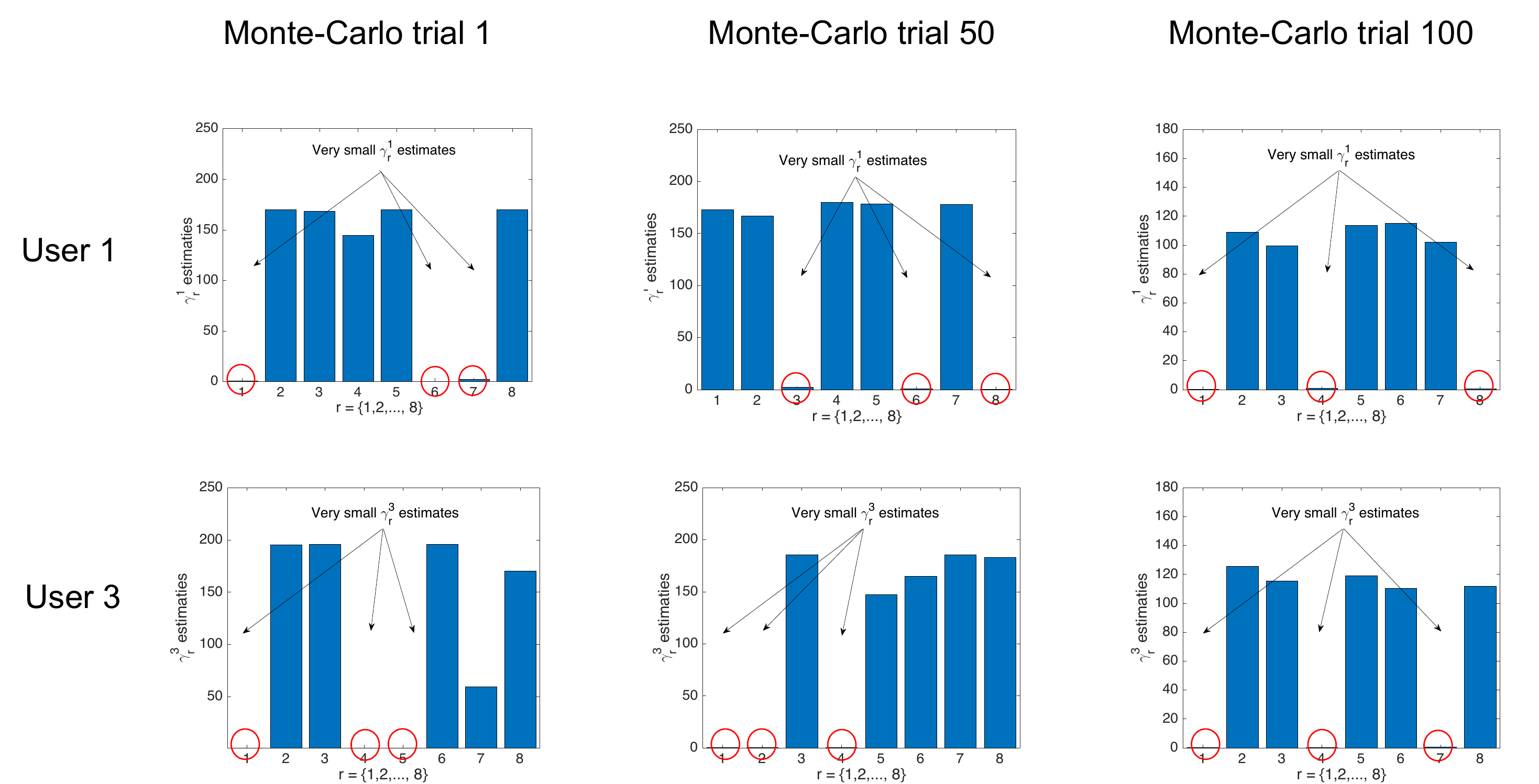}
\caption{The estimates of $\{\gamma_r^n\}_{r=1}^{\bar{R}^n}$ for user 1 and user 3  in different Monte-Carlo trials ($\bar{R}^n = 8$, $R^n =3$, $L = 10$, SNR = 20 dB).}
\label{fig_topology}
\end{figure*}

\begin{figure*}[!t]
\centering
\includegraphics[width= 7 in]{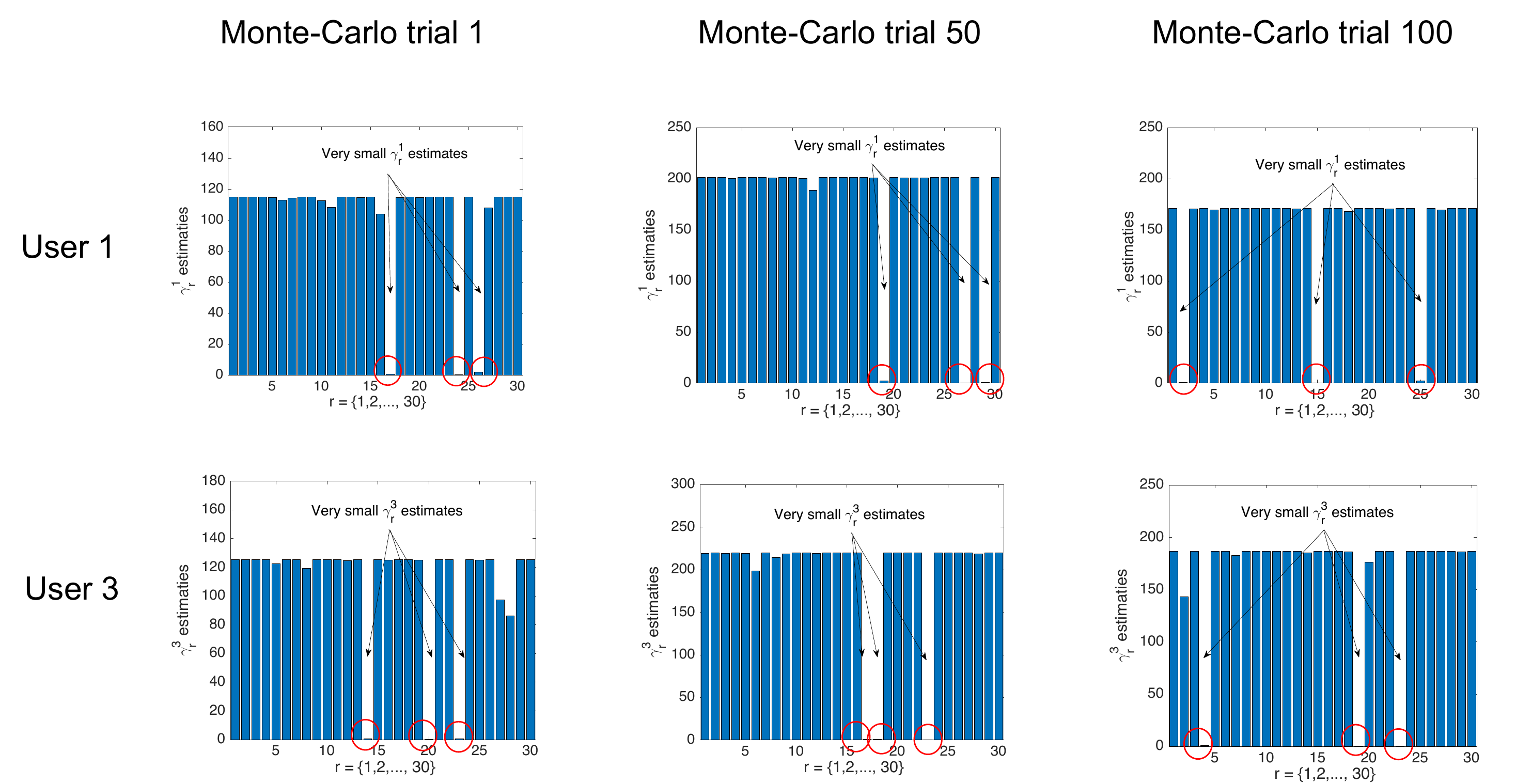}
\caption {The estimates of $\{\gamma_r^n\}_{r=1}^{\bar{R}^n}$ for user 1 and user 3  in different Monte-Carlo trials  ($\bar{R}^n = 30$, $R^n =3$, $L = 10$, SNR = 20 dB).}
\label{fig_topology}
\end{figure*}

As discussed in Section IV. E, the estimation results of $\{\gamma_r^n\}_{r=1}^{\bar{R}^n}$ under different ${\bar R}^n$s  determine the channel model complexity learning performance of the proposed method. Since  $\{\gamma_r^n\}_{r=1}^{\bar{R}^n}$ in different Monte-Carlo trials are possibly  with different sparsity patterns (i.e., the very small values might appear in different subscripts $r$), averaging them over Monte-Carlo trials is not informative. Therefore,  in Figure 5 and Figure 6, we present the estimation results of $\{\gamma_r^n\}_{r=1}^{\bar{R}^n}$ for user 1 and user 3 in three independent Monte-Carlo trials under ${\bar R}^n = 8$ and ${\bar R}^n = 30$ respectively.  From these two figures, it can be seen that under both ${\bar R}^n = 8$ and ${\bar R}^n = 30$, only three $\gamma_r^n$s were estimated to be very small, indicating that there are three significant channel paths for each user. Since the exact path number $R_n =3$, it shows that the proposed method can accurately recover the channel model complexity and thus avoid the overfitting.

\subsection{Channel Estimation Performance}

To assess the channel estimation performance at different SNRs, the MSEs of different algorithm are shown in Figure 7. From Figure 7, it is obvious that the three tensor-aided methods (VI-$\bar{R}^n$, BCD-$\bar{R}^n$, and BCD-$R^n$) achieve much more accurate channel estimation than the LS method, due to the exploitation of tensor structures in the adopted angular channel model. It can be also observed that the genie-aided BCD-$R^n$ method achieves the lowest MSE for a wide range of SNRs, since it fits the channel coefficients into the observation data assuming the accurate channel model complexity, which however is not available in practice. With a wrong guess of the path numbers, the MSEs of the BCD-$\bar{R}^n$ method are much higher than those of the BCD-$R^n$ algorithm, due to the overfitting of noises. In contrast, although the proposed VI-$\bar{R}^n$ algorithm is also with a wrong guess of the path numbers, its MSEs are nearly the same as those of the genie-aided BCD-$R^n$ method. This shows the effectiveness of the Bayesian method in automatic model complexity control and overfitting avoidance.  

\begin{figure}[!t]
\centering
\includegraphics[width=3.5 in]{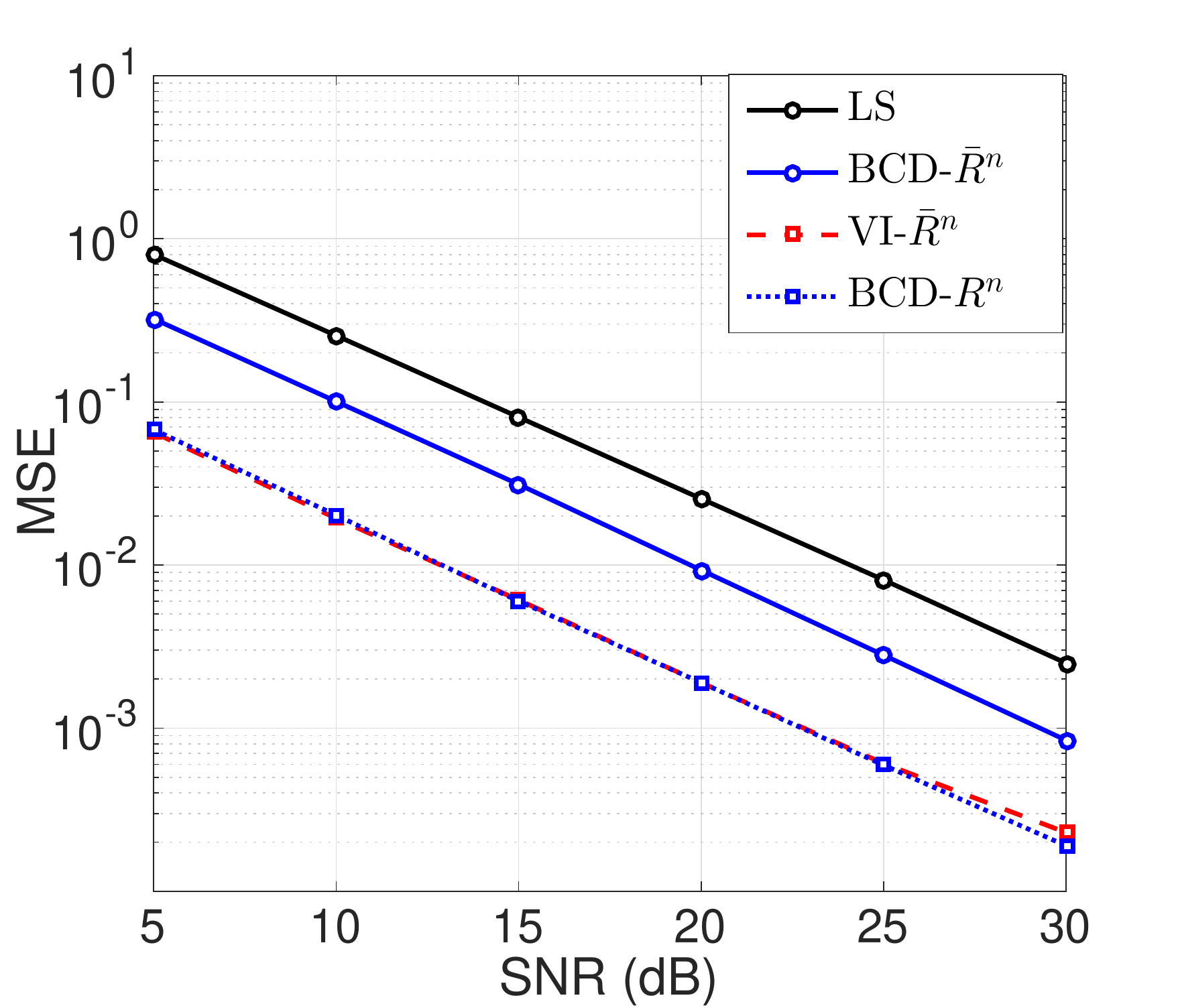}
\caption{Performance of channel estimation versus SNRs ($\bar{R}^n$ = 8, $R^n = 3, L = 10$).}
\label{fig_topology}
\end{figure}

On the other hand, we present the running time of the three iterative tensor-aided channel estimation algorithms  (VI-$\bar{R}^n$, BCD-$\bar{R}^n$, and BCD-$R^n$) in Table II.  From Table II, it can be observed that  the proposed algorithm is with comparable running time to that of the  BCD-$\bar{R}^n$ approach, which validates the complexity analysis in Section IV. E. Notice that these two approaches cost much more time than the genie-aided BCD-$R^n$ algorithm,  since they need to update over-determined model parameters.

\begin{table}[!h]
\caption{Running time (second) of different channel estimation algorithms ($\bar{R}^n = 8$, ${R}^n =3$, $L = 10$)}
\centering 
\begin{tabular}{@{}|c|c|c|c|@{}}
\toprule
SNR   & BCD-$R^n$  & BCD-$\bar{R}^n$ & VI-$\bar{R}^n$ \\ \midrule
10 dB & 1.7942 & 6.5550   & 6.6890  \\ \midrule
20 dB & 1.5422 & 4.1581   & 4.9155  \\ \bottomrule
\end{tabular}
\end{table}

To see how the model complexity of channels affects different algorithms, under SNR = 20 dB, the MSEs of channel estimations versus different path numbers in the channel model are presented in Figure 8. In previous simulation studies,  $R^n = 3, \forall n$ is considered. Here we further consider different path number values $R^n = \{2,3,4,5,6\}$, each of which indicates different channel model complexities. With a higher $R^n $, there are more unknown channel coefficients. Then, it is expected that the channel estimation performance would degrade given the same amount of the observation data. This conjecture has been validated by Figure 8, in which the MSEs indeed increase as $R^n$ increases. On the other hand, it can be seen that the proposed  VI-$\bar{R}^n$ algorithm achieves indistinguishable performances as those of the genie-aided BCD-$R^n$ method. This shows that the proposed algorithm can learn a wide range of  model complexities and then effectively shrink redundant channel model parameters for overfitting avoidance.

\begin{figure}[!t]
\centering
\includegraphics[width=3.5 in]{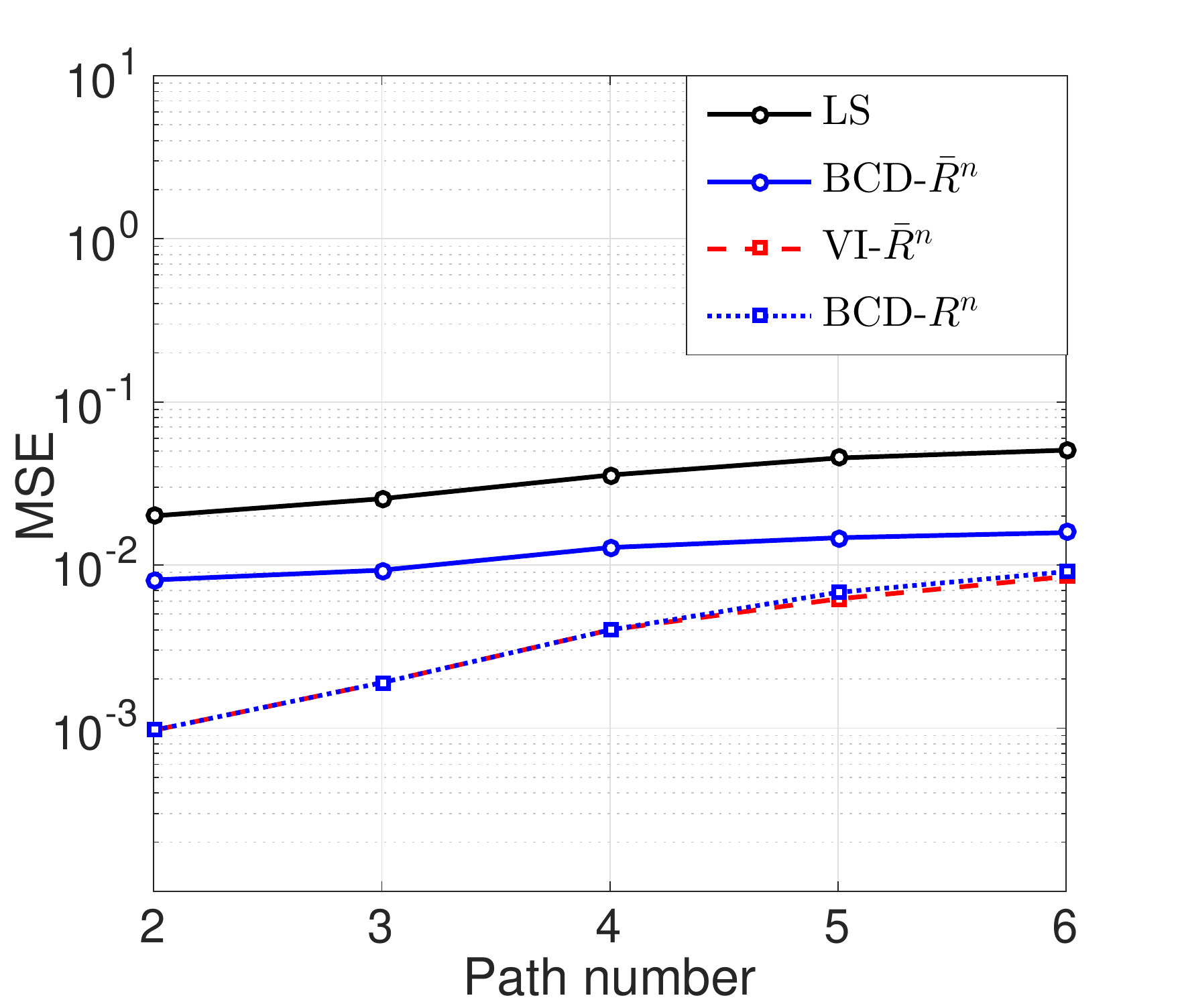}
\caption{Performance of channel estimation versus different path numbers (SNR = 20 dB, $\bar{R}^n = 8, L = 10$).}
\label{fig_topology}
\end{figure}

\begin{figure}[!t]
\centering
\includegraphics[width=3.5 in]{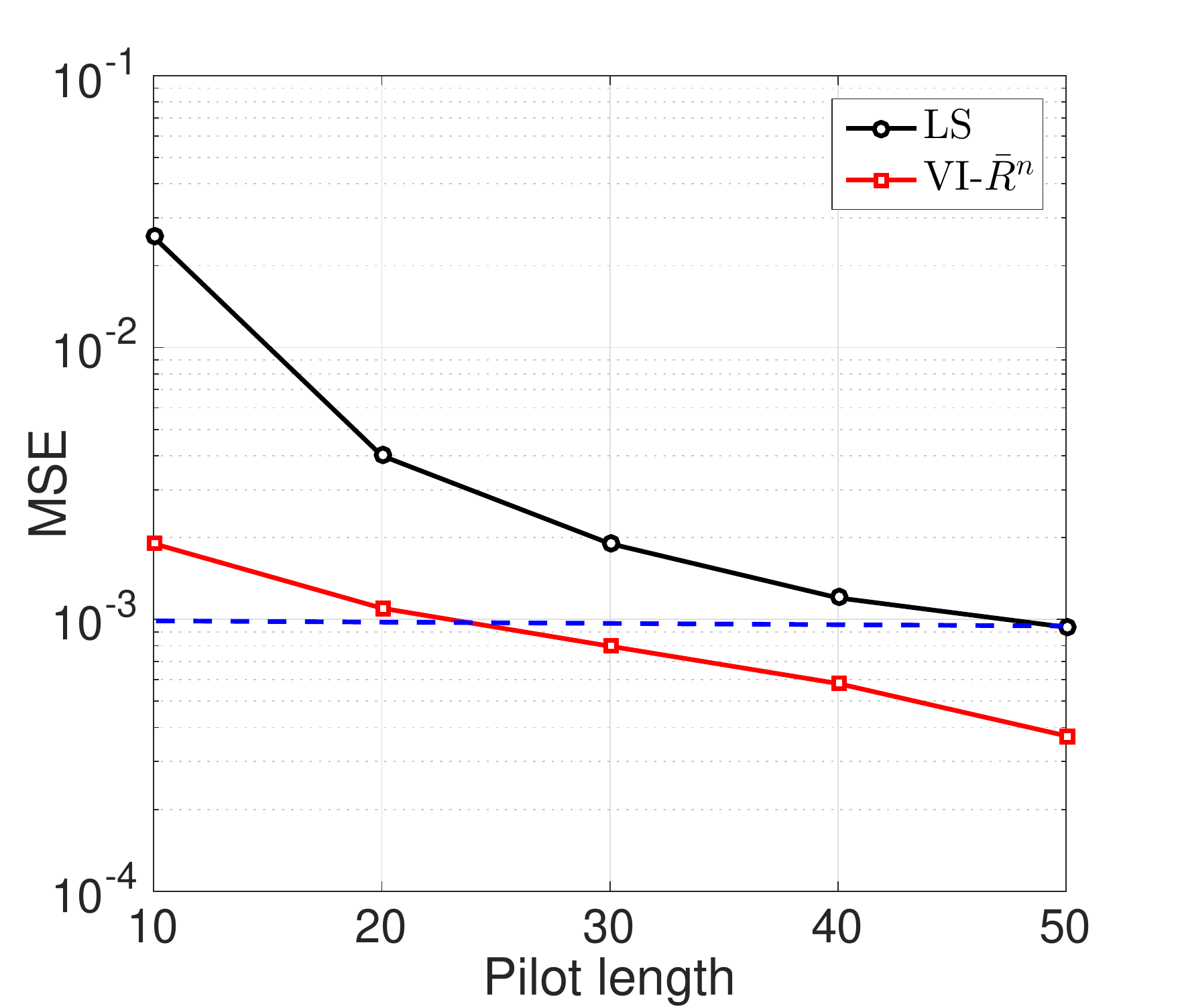}
\caption{Performance of channel estimation versus different pilot lengths (SNR = 20 dB, $\bar{R}^n = 8, R^n = 3$).}
\label{fig_topology}
\end{figure}

Finally, in Figure 9, we show how the proposed VI-$\bar{R}^n$ algorithm saves the pilot resources  for channel estimation, compared to the standard LS method. From Figure 9, it is clear that given the MSE $10^{-3}$, the proposed  VI-$\bar{R}^n$ algorithm needs only around $20$ pilot signals while the LS method requires about $50$ pilot signals. The gain comes from both the tensor structure exploitation of channel model and the Bayesian philosophy in overfitting avoidance.

\section{Conclusions and Future Research}
In this paper, the multi-user channel estimation problem for 3D massive MIMO communications was investigated through the lens of  Bayesian tensor methods. The channel estimation problem was firstly recasted as a factor matrix learning problem for a non-standard tensor decomposition model, which requires a novel learning algorithm design with an integrated feature of overfitting avoidance. To achieve this goal, a tuning-free channel estimation algorithm was proposed in this paper under the framework of Bayesian modelling and inference. Numerical studies have shown the excellent channel estimation performance of the proposed method in terms of both accuracy and overfitting avoidance. 

In future research,  the exploitation of the shift-invariance property [21], \cite{A5} in cubic antenna array at the base station might give a new tensor model for massive MIMO communications, which also needs novel tuning-free channel estimation algorithm  design. \emph{We believe that the integration of tensor model, array signal processing and Bayesian method will bring us closer to the era of ``Joint Model-and-Data-Driven Wireless Communications”.}

\appendices

\section{Uniqueness Property of Tensor CPD}
In [40], a sufficient condition for the uniqueness of tensor CPD is stated as follows. 

\noindent {\bf Uniqueness condition for CPD [40].} \emph{If $\llbracket \boldsymbol A^{(1)},  \boldsymbol A^{(2)}, ...,  \boldsymbol A^{(N)}\rrbracket $ =  $ \llbracket \boldsymbol \Xi^{(1)}, \boldsymbol \Xi^{(2)}, ..., \boldsymbol \Xi^{(N)} \rrbracket $, and $\sum_{n=1}^{N} k_n \geq 2L + (N-1)$ where $k_i$ denotes the k-rank of matrix $\boldsymbol A^{(i)}$ and $L$ is the tensor rank. Then the following equations hold:  $\boldsymbol \Xi^{(1)} = \boldsymbol A^{(1)} \boldsymbol \Delta \boldsymbol \Lambda^{(1)}$, $\boldsymbol \Xi^{(2)} =  \boldsymbol A^{(2)} \boldsymbol \Delta \boldsymbol \Lambda^{(2)}$, ...,  $\boldsymbol \Xi^{(N)} =\boldsymbol A^{(N)} \boldsymbol \Delta \boldsymbol \Lambda^{(N)} $  where $\boldsymbol  \Delta$ is a permutation matrix and  diagonal matrix $\boldsymbol \Lambda^{(n)}$ satisfies $\prod_{n=1}^N \boldsymbol \Lambda^{(n)} = \boldsymbol I_{L}$ .}
 
Thus tensor CPD is unique up to trivial scaling and permutation ambiguities.

\section{Complicated Coupling after Expanding the Frobenius norm in (8)}
Since $|| \mathcal X ||_F^2 = || \mathcal X(k)||_F^2$ where $\mathcal X(k)$ is the $k^{th}$ unfolding matrix of the tensor $\mathcal X$, after using the unfolding property of tensor CPD [40],  problem (8) is equivalent to 
\begin{align}
&\min_{\{  \{\boldsymbol \Xi^{(k),n} \}_{k=1}^3 \}_{n=1}^N}  \sum_{l=1}^L  \bigparallel  
 \mathcal Y_l (k) -   \nonumber \\
& ~~~~~~~~~~~~~~\sum_{n=1}^N s_n(l) \boldsymbol  \Xi^{(k),n} \left( \mathop \diamond  \limits_{j=1,j\ne k}^3  \left[\boldsymbol \Xi^{(j),n}\right]\right)^T{\bigparallel}_F^2. 
\end{align}
Further using the fact $||\mathbf X ||_F^2 = \mathrm{Tr}(\mathbf X \mathbf X^H)$ to expand the Frobenius norm in (36), we have the following problem:
\begin{align}
&\min_{\{  \{\boldsymbol \Xi^{(k),n} \}_{k=1}^3 \}_{n=1}^N}  \sum_{l=1}^L  \mathrm{Tr} \Bigg( \Big[
 \mathcal Y_l (k) -  \sum_{n=1}^N s_n(l) \boldsymbol  \Xi^{(k),n} \nonumber \\
 &\times \left( \mathop \diamond  \limits_{j=1,j\ne k}^3  \left[\boldsymbol \Xi^{(j),n}\right]\right)^T \Big] \Big[  \mathcal Y_l (k) -  \sum_{n=1}^N s_n(l) \boldsymbol  \Xi^{(k),n} \nonumber \\
&~~~~~~~~~~~~~~~~~~~~~~~~~~ \times  \left( \mathop \diamond  \limits_{j=1,j\ne k}^3  \left[\boldsymbol \Xi^{(j),n}\right]\right)^T \Big]^H \Bigg). 
\end{align}
\begin{figure*}[!t]
\normalsize
\begin{align}
&\min_{\{  \{\boldsymbol \Xi^{(k),n} \}_{k=1}^3 \}_{n=1}^N}  \sum_{l=1}^L  \mathrm{Tr} \Bigg( 
\mathcal Y_l (k)  \mathcal Y_l (k)^H -  2\mathfrak{Re} \Big( \mathcal Y_l(k) \sum_{n=1}^N s_n^*(l) \left( \mathop \diamond  \limits_{j=1,j\ne k}^3  \left[\boldsymbol \Xi^{(j),n}\right]\right)^* \big[\boldsymbol  \Xi^{(k),n}  \big]^H \Big)  \nonumber \\
&+  \underbrace{ \Big[\sum_{n=1}^N s_n(l) \boldsymbol  \Xi^{(k),n} \left( \mathop \diamond  \limits_{j=1,j\ne k}^3  \left[\boldsymbol \Xi^{(j),n}\right]\right)^T \Big] \Big[
  \sum_{n=1}^N s_n(l) \boldsymbol  \Xi^{(k),n} \left( \mathop \diamond  \limits_{j=1,j\ne k}^3  \left[\boldsymbol \Xi^{(j),n}\right]\right)^T \Big]^H}_{\mathfrak {t}} \Bigg).
\end{align}
\hrulefill
\vspace*{4pt}
\end{figure*} 

After manipulating algebras, problem (37) becomes (38) at the top of the next page. {In the term $\mathfrak t$, it is clear that the product of two summation terms, i.e.,  
\begin{align}
& \Big[\sum_{n=1}^N s_n(l) \boldsymbol  \Xi^{(k),n} \left( \mathop \diamond  \limits_{j=1,j\ne k}^3  \left[\boldsymbol \Xi^{(j),n}\right]\right)^T\Big] \nonumber \\
&\times  \Big[\sum_{n=1}^N s_n(l) \boldsymbol  \Xi^{(k),n} \left( \mathop \diamond  \limits_{j=1,j\ne k}^3  \left[\boldsymbol \Xi^{(j),n}\right]\right)^T\Big]^H
 \end{align}
will result in complicated coupling among the factor matrices $\{ \{ \boldsymbol \Xi^{(k),n}\}_{k=1}^3\}_{n=1}^N$.} Although problem (38) seems complicated, if we only optimize a single factor matrix $ \boldsymbol \Xi^{(k),n}$ while fixing other variables, problem (38) will become problem (9) in Section III, which is a convex problem and can be easily solved.

\section{The Derivations of The Optimal Variational Pdfs in Table I}
After substituting (19) into (18) and only keep terms relevant to $\boldsymbol \Xi^{(k),n}$, we have 
\begin{align}
Q^\dagger \left(  \boldsymbol \Xi^{(k),n}\right) &\propto \exp \Bigg \{  \mathbb E\Bigg[ - \beta \sum_{l=1}^L \bigparallel  \mathcal Y_l -  s_n(l)   \llbracket \boldsymbol \Xi^{(1),n}, \boldsymbol \Xi^{(2),n}, \nonumber\\
& \boldsymbol \Xi^{(3),n}\rrbracket    - \sum_{p=1, p\neq n}^N s_p(l)  \llbracket \boldsymbol \Xi^{(1),p}, \boldsymbol \Xi^{(2),p}, \boldsymbol \Xi^{(3),p}  \rrbracket {\bigparallel}_F^2 \nonumber\\
&  - \mathrm{Tr} \left( \boldsymbol \Xi^{(k),n} \boldsymbol \Gamma^n  \left[\boldsymbol \Xi^{(k),n}\right]^H \right)\Bigg] \Bigg\}.
\end{align}
Then, we utilize the result $\parallel \boldsymbol A \parallel_F^2 = \mathrm{Tr}(\boldsymbol  A \boldsymbol A^H)$ to expand the Frobenius norm. After a series of  algebra manipulations, $Q^\dagger \left(  \boldsymbol \Xi^{(k),n}\right)$ can be organized to be 
\begin{align}
& Q^\dagger \left(  \boldsymbol \Xi^{(k),n}\right) \nonumber \\
& \propto \exp \Bigg \{  \mathbb E\Bigg[ \mathrm{Tr}\Bigg( -\boldsymbol \Xi^{(k),n}  \Bigg(  \sum_{l=1}^L |s_n(l)|^2\beta \left(\mathop \diamond  \limits_{j=1,j\ne k}^3 \boldsymbol \Xi^{(j),n}\right)^T \nonumber\\
& \times \left(\mathop \diamond  \limits_{j=1,j\ne k}^3 \boldsymbol \Xi^{(j),n}\right)^*  +  \boldsymbol \Gamma^n 
\Bigg)
\left[ \boldsymbol \Xi^{(k),n} \right]^H + 2\mathfrak{Re} \Bigg(\boldsymbol \Xi^{(k),n} \nonumber \\
& \times \sum_{l=1}^L s_n(l) \beta  \left( \mathop \diamond  \limits_{j=1,j\ne k}^3 \boldsymbol \Xi^{(j),n}\right)^T
\Bigg(  \mathcal Y_l (k) \!-\!  \sum_{p=1, p\neq n}^N s_p(l) \nonumber \\
&\times \Big \llbracket \boldsymbol \Xi^{(1),p}, \boldsymbol \Xi^{(2),p}, \boldsymbol \Xi^{(3),p} \Big \rrbracket (k)\Bigg) \Bigg)\Bigg)\Bigg] \Bigg\}.
\end{align}
After distributing the expectations and comparing the functional form of (41) to that of circularly-symmetric complex matrix normal distribution \cite{MND}, it can be concluded that $Q^\dagger \left(  \boldsymbol \Xi^{(k),n}\right) = \mathcal{CMN}(\boldsymbol \Xi^{(k),n} |\boldsymbol M^{(k),n}, \boldsymbol I_{I_k}, \boldsymbol \Sigma^{(k),n})$ with its mean $\boldsymbol M^{(k),n}$ and covariance matrix $\boldsymbol \Sigma^{(k),n}$ being defined in (20) and (21). 

Similarly, after substituting (19) and (18), and fixing all the variables other than $\{\{\gamma_r^n\}_{r=1}^{\bar{R}^n}\}_{n=1}^N$, we have
\begin{align}
& Q^\dagger \left( \{\{\gamma_r^n\}_{r=1}^{\bar{R}^n}\}_{n=1}^N \right) \propto  \exp \Bigg \{\mathbb E\Bigg[ \sum_{n=1}^N \sum_{k=1}^3 -\mathrm{Tr} \Big( \boldsymbol \Xi^{(k),n}  \boldsymbol \Gamma^n  \nonumber \\
&\left[\boldsymbol \Xi^{(k),n}\right]^H \Big) +I_k \sum_{r=1}^{\bar{R}^n}   \ln \gamma_r^n + (\epsilon-1)\ln \gamma_r^n -\epsilon \gamma_r^n \Bigg] \Bigg\}.
\end{align}
Using the fact that $\mathrm{Tr} \Big( \boldsymbol \Xi^{(k),n}  \boldsymbol \Gamma^n \left[\boldsymbol \Xi^{(k),n}\right]^H \Big) = \sum_{r=1}^{\bar{R}^n} \gamma_r^n  \left[\boldsymbol \Xi^{(k),n}\right]_{:,r}^H \left[\boldsymbol \Xi^{(k),n}\right]_{:,r}  $, it can be shown that 
\begin{align}
& Q^\dagger \left( \{\{\gamma_r^n\}_{r=1}^{\bar{R}^n}\}_{n=1}^N \right) \propto   \prod_{n=1}^N  \prod_{r=1}^{\bar{R}^n} 
\exp \Bigg \{\mathbb E\Bigg[ -\gamma_r^n \sum_{k=1}^3  \Big( \left[\boldsymbol \Xi^{(k),n}\right]_{:,r}^H  \nonumber \\
& \left[\boldsymbol \Xi^{(k),n}\right]_{:,r} + I_k  \ln \gamma_r^n \Big)+ (\epsilon-1)\ln \gamma_r^n -\epsilon \gamma_r^n \Bigg] \Bigg\}.
\end{align}
It is easy to conclude that $ Q^\dagger \left( \{ \{ \gamma_r^n\}_{r=1}^{\bar{R}^n} \}_{n=1}^N \right)  = \prod_{n=1}^N \prod_{r=1}^{\bar{R}^n} Q^\dagger(\gamma_r^n)$, where 
\begin{align}
& Q^\dagger(\gamma_r^n)  \propto \exp \Bigg\{ \left( \sum_{k=1}^3 I_k + \epsilon - 1\right)\ln \gamma_r^n \nonumber\\
 & - \gamma_r^n \left( \epsilon +  \sum_{k=1}^3\mathbb E \left[ \left[\boldsymbol \Xi^{(k),n}\right]_{:,r}^H \left[\boldsymbol \Xi^{(k),n}\right]_{:,r} \right] \right) \Bigg\}.
\end{align}
By comparing (44) to the functional form of gamma distribution, we have $Q^\dagger \left( \gamma_r^n\right)  =  \mathrm{gamma}(\gamma_r^n | a_r^n, b_r^n)$, where $a_r^n, b_r^n$ is defined by (23) and (24) respectively.

Finally, we use (18) and (19) again to derive the optimal variational pdf $Q^\dagger \left( \beta\right)$. It can be shown that 
\begin{align}
&Q^\dagger \left( \beta\right) \propto \exp \Bigg\{ \left(\prod_{k=1}^3 I_k L + \epsilon -1 \right)\ln \beta  \nonumber\\
& -\beta \Bigg( \epsilon + \sum_{l=1}^L  \mathbb E \Bigg[ \bigparallel  \mathcal Y_l -  \sum_{n=1}^N s_n(l) \llbracket \boldsymbol \Xi^{(1),n}, \boldsymbol \Xi^{(2),n},\boldsymbol \Xi^{(3),n} \rrbracket  {\bigparallel}_F^2\Bigg] \Bigg)
\Bigg\}.
\end{align}
After comparing (45) to the functional form of gamma distribution, it is easy to identify $Q^\dagger \left( \beta\right)  = \mathrm{gamma}(\beta | c, d)$, where $c$ and $d$ are expressed in (25) and (26) respectively. 


\section{Expectation Computation for (26)}
In (26), computing the expectation $ \mathbb E \left[ \bigparallel  \mathcal Y_l -  \sum_{n=1}^N s_n(l) \llbracket \boldsymbol \Xi^{(1),n}, \boldsymbol \Xi^{(2),n}, \boldsymbol \Xi^{(3),n} \rrbracket  {\bigparallel}_F^2\right] $ is quite complicated. We use the result $\parallel \boldsymbol A \parallel_F^2 = \mathrm{Tr}(\boldsymbol  A \boldsymbol A^H)$ and the tensor unfolding property \cite{tensor1} to expand the Frobenius norm:
\begin{align}
&\mathbb E \left[ \bigparallel  \mathcal Y_l -  \sum_{n=1}^N s_n(l) \llbracket \boldsymbol \Xi^{(1),n}, \boldsymbol \Xi^{(2),n}, \boldsymbol \Xi^{(3),n} \rrbracket  {\bigparallel}_F^2\right]   \nonumber \\
& = \mathbb E \Bigg[ \mathrm{Tr} \Bigg( \mathcal Y_l (1) \mathcal Y_l (1)^H - 2\mathfrak{Re}\Bigg(\mathcal Y_l(1) \sum_{n=1}^N s_n(l)^*   \nonumber\\
& \left(\mathop \diamond  \limits_{j=2}^3 \boldsymbol \Xi^{(j),n}\right)^*\left[\boldsymbol \Xi^{(1),n}\right]^H \Bigg)+  \mathcal G_l (1) \mathcal  G_l(1)^H\Bigg)\Bigg],
\end{align}
where 
\begin{align}
\mathcal  G_l =   \sum_{n=1}^N s_n(l) \llbracket \boldsymbol \Xi^{(1),n}, \boldsymbol \Xi^{(2),n}, \boldsymbol \Xi^{(3),n} \rrbracket.  
\end{align}
After distributing the expectations, the most complicated term is $\mathbb E \left[\mathcal G_l(1) \mathcal G_1(1)^H\right]$. Using the tensor unfolding property  \cite{tensor1} again, we have 
\begin{align}
&\mathbb E \left[\mathcal G_l(1) \mathcal G_1(1)^H\right] \nonumber \\
& = \mathbb E \Bigg[ \mathrm{Tr} \Bigg(\Bigg[ \sum_{n=1}^N s_n(l)  \boldsymbol \Xi^{(1),n} \left(\mathop \diamond  \limits_{j=2}^3 \boldsymbol \Xi^{(j),n}\right)^T \Bigg] \nonumber \\
& ~~\times \Bigg[ \sum_{n=1}^N s_n(l)  \boldsymbol \Xi^{(1),n} \left(\mathop \diamond  \limits_{j=2}^3 \boldsymbol \Xi^{(j),n}\right)^T \Bigg]^H \Bigg) \Bigg] \nonumber \\
& = \mathrm{Tr} \Bigg( \sum_{n=1}^N \sum_{p=1}^N  s_n(l) s_p(l)^*  \mathbb E \Bigg[ \boldsymbol \Xi^{(1),n}   \left(\mathop \diamond  \limits_{j=2}^3 \boldsymbol \Xi^{(j),n}\right)^T \nonumber \\
 & ~~~~~~\times \left(\mathop \diamond  \limits_{j=2}^3 \boldsymbol \Xi^{(j),p}\right)^* \left[\boldsymbol \Xi^{(1),p}  \right]^H  \Bigg]\Bigg).
\end{align}
Further using the results in (27) and (28), we have 
\begin{align}
&\mathbb E \left[\mathcal G_l(1) \mathcal G_1(1)^H\right] \nonumber \\
& =\mathrm{Tr} \Bigg( \sum_{n=1}^N \sum_{p=1, p \neq n}^N  s_n(l) s_p(l)^* \boldsymbol M^{(1),n} \left(\mathop \diamond  \limits_{j=2}^3 \boldsymbol M^{(j),n}\right)^T \nonumber \\
& \times \left(\mathop \diamond  \limits_{j=2}^3 \boldsymbol M^{(j),p}\right)^* \left[\boldsymbol M^{(1),p}\right]^H \Bigg)\nonumber\\
& + \mathrm{Tr} \Bigg( \sum_{n=1}^N |s_n(l)|^2 \left [  \left[\boldsymbol M^{(1),n}\right]^H  \boldsymbol M^{(1),n} + I_1 \boldsymbol \Sigma^{(1),n} \right]  \nonumber \\
&  \mathop \odot \limits_{k=2}^3 \left [  \left[\boldsymbol M^{(k),n}\right]^H  \boldsymbol M^{(k),n} + I_k \boldsymbol \Sigma^{(k),n} \right]^* \Bigg).
\end{align}
After putting (49) into (46), the result of (29) can be obtained.

\end{document}